\documentclass[12pt]{article}
\usepackage{graphicx,xspace,amsmath,amssymb,epsfig,aeguill,aecompl}
\graphicspath{{Eps/}} 
\input{Input/alphabet}   
\input{Input/abrege}     
\def\th{{\mathrm{th}}}                

\newsavebox{\fminibox}
\newlength{\fminilength}
\newenvironment{fminipage}[1][\linewidth]
  {\setlength{\fminilength}{#1}
   \begin{lrbox}{\fminibox}\begin{minipage}{\fminilength}}
  {\end{minipage}\end{lrbox}\noindent\fbox{\usebox{\fminibox}}}

  \def\+{^\dagger}

\def\nequiv{\not\kern-.05em\equiv}
\def\egal{\kern-.5em=\kern-.5em}        
\def\propt{\kern-.2em\propto\kern-.2em} 

\def\argmax{\mathop{\mathrm{arg\,max}}} 
\def\argmin{\mathop{\mathrm{arg\,min}}} 
\def\intdouble{\int\kern-0.3em\int}
\def\inttriple{\int\kern-0.3em\int\kern-0.3em\int}

\def\rond#1{\overset{\kern-0.33em~_\circ}{#1}}
\def\rondit[#1]#2{\overset{\kern#1~_\circ}{#2}}

\def\babs{\begin{abstract}}             \def\eabs{\end{abstract}}
\def\barr{\begin{array}}                \def\earr{\end{array}}
\def\bcc{\begin{center}}                \def\ecc{\end{center}}

\def\bdes{\begin{description}}          \def\edes{\end{description}}
\def\bdoc{
\begin{document}}             \def\edoc{\end{document}}
\def\ben{\begin{enumerate}}             \def\een{\end{enumerate}}
\def\beqn{\begin{eqnarray}}             \def\eeqn{\end{eqnarray}}
\def\beqnl#1{\beqn\label{#1}}           \def\eeqnl#1{\label{#1}\eeqn}
\def\beqnx{\begin{eqnarray*}}           \def\eeqnx{\end{eqnarray*}}
\def\bseqn{\begin{subeqnarray}}         \def\eseqn{\end{subeqnarray}}
\def\beq#1\eeq{\begin{equation}#1\end{equation}}
\def\bal#1\eal{\begin{align}#1\end{align}}
\def\balx#1\ealx{\begin{align*}#1\end{align*}}
\def\beqx{$$}                           \def\eeqx{$$}
\def\bfig{\protect\begin{figure}}       \def\efig{\protect\end{figure}}
\def\bfigx{\protect\begin{figure*}}     \def\efigx{\protect\end{figure*}}
\def\bfigt{\protect\begin{figurette}}   \def\efigt{\protect\end{figurette}}
\def\bfl{\begin{flushleft}}             \def\efl{\end{flushleft}}
\def\bfr{\begin{flushright}}            \def\efr{\end{flushright}}
\def\bit{\begin{itemize}}               \def\eit{\end{itemize}}
\def\bmi{\begin{minipage}}              \def\emi{\end{minipage}}
\def\bfmi{\begin{fminipage}}            \def\efmi{\end{fminipage}}
\def\bpic{\begin{picture}}              \def\epic{\end{picture}}
\def\bqu{\begin{quote}}                 \def\equ{\end{quote}}
\def\bqun{\begin{quotation}}            \def\equn{\end{quotation}}
\def\bsl{\begin{slide}}                 \def\esl{\end{slide}}
\def\btabb{\begin{tabbing}}             \def\etabb{\end{tabbing}}
\def\btabl{\begin{table}}               \def\etabl{\end{table}}
\def\btablx{\begin{table*}}             \def\etablx{\end{table*}}
\def\btab{\begin{tabular}} 
\def\btabu{\begin{tabular}}             \def\etabu{\end{tabular}}
\def\btabx{\begin{tabular*}}            \def\etabx{\end{tabular*}}
\def\bbib{\begin{thebibliography}{999}} \def\ebib{\end{thebibliography}}
\def\bver{\begin{verbatim}}             \def\ever{\end{verbatim}}
\def\bca{\begin{cases}}                  \def\eca{\end{cases}}
   
\def\bm#1{\mbox{\bf #1}}

\def\d#1{\,\hbox{d}#1}
\def\bs{\bar{s}}
\def\us{\underline{s}}
\def\fmin{f_{\mbox{min}}}
\def\fmax{f_{\mbox{max}}}
\def\expf#1{\mbox{exp}\left\{#1\right\}}
\def\argmin#1#2{\mbox{arg}\min_{#1}\left\{#2\right\}}
\def\argmax#1#2{\mbox{arg}\max_{#1}\left\{#2\right\}}

\def\lra{\longrightarrow}
\def\fh{\widehat{f}}
\def\fbh{\widehat{\fb}}
\def\Rbh{\widehat{\Rb}}
\def\Sigmabh{\widehat{\Sigmab}}

\def\rbh{\widehat{\rb}}
\def\qbh{\widehat{\qb}}
\def\xbh{\widehat{\xb}}
\def\tbh{\widehat{\tb}}
\def\thetabbh{\widehat{\thetabb}}

\def\disp{\displaystyle}
\def\vsm{\vspace*{-12pt}}
\def\hsm{\hspace*{-3em}}

\def\pmata#1#2{\left(\barr{c} #1 \\ #2 \earr\right)}
\def\pmatb#1#2#3#4{\left(\barr{cc} #1 & #2 \\ #3 & #4 \earr\right)}

\def\th{\widehat{t}}
\def\xh{\widehat{x}}
\def\lambdah{\widehat{\lambda}}
\def\muh{\widehat{\mu}}
\def\sigmah{\widehat{\sigma}}
\def\alphah{\widehat{\alpha}}
\def\betah{\widehat{\beta}}
\def\tbh{\widehat{\tb}}
\def\mubh{\widehat{\mub}}
\def\sigmabh{\widehat{\sigmab}}
\def\alphabh{\widehat{\alphab}}


\title{A Hidden Markov model for Bayesian data fusion of multivariate signals}

\author{Olivier F\'eron and Ali Mohammad-Djafari\\[.4cm]
  Laboratoire des Signaux et Syst\`emes,\\  
  Unit\'e mixte de recherche 8506 (CNRS-Sup\'elec-UPS) \\  
  Sup\'elec, Plateau de Moulon, 91192 Gif-sur-Yvette, France\\ 
  emails = {feron@lss.supelec.fr} \quad {djafari@lss.supelec.fr}
}

\date{}

\bdoc

\maketitle

\begin{abstract}
\noindent
In this work we propose a Bayesian framework for data fusion of multivariate signals which arises in imaging systems. More specifically, we consider the case where we have observed two images of the same object through two different imaging processes. The objective of this work is then to propose a coherent approach to combine these data sets to obtain a segmented image which can be considered as the fusion result of these two images. \\
The proposed approach is based on a Hidden Markov Modeling (HMM) of the images with common segmentation, or equivalently, with common hidden classification label variables which is modeled by the Potts Markov Random Field. We propose then an appropriate Markov Chain Monte Carlo (MCMC) algorithm to implement the method and show some simulation results and applications. \\

\noindent
{\bf key words :} \\
Data fusion, Classification and segmentation of Images, HMM, MCMC, Gibbs Algorithm. \\
\end{abstract}

\section{Introduction}

\noindent
Data fusion and multi-source information has become a very active area of research in many domains : industrial nondestructive testing and evaluation (\cite{Gautier95b}), medical imaging \cite{Boyd94,Boyd95,Matsopoulos}, industrial inspection (\cite{Bass00})(quality control and condition monitoring) and security systems in general. \\
\noindent
In all these areas, the main problem is how to combine the information contents of different sets of multivariate data $g_i(r)$. When the data set $g_i(r)$ is an image we have $r \in \Rb^2$, and the problem becomes how to combine and represent their fusion. Very often the data sets $g_i$ do not represent the same quantities. For example in medical imaging we have 2D radiographic data $g_1$ and echographical data $g_2$ which are related to different properties $f_1$ and $f_2$ of the body under examination by
\beq
g_i(r)=[H_if_i](r) + \varepsilon_i(r)
\eeq
\noindent
where $H_i$ are the operator functionnal of the measuring systems. We may note that estimating $f_i$ given each set of data $g_i$ is an inverse problem by itself which is often an ill-posed problem even if we may know perfectly the operator $H_i$. So very often people use the two data sets separately to obtain two images $f_1$ and $f_2$ and then they try to make a data fusion. We think it is possible to do a better job if we define more precisely what we mean by data fusion of two images $f_1$ and $f_2$ and if we try to use the data $g_1$ and $g_2$ to estimate directly not only $f_1$ and $f_2$ but also the common feature of them which we present by a third image $z$. \\
\begin{figure}[h]
\centering
\begin{tabular}{cccc}
	\mbox{\epsfig{file=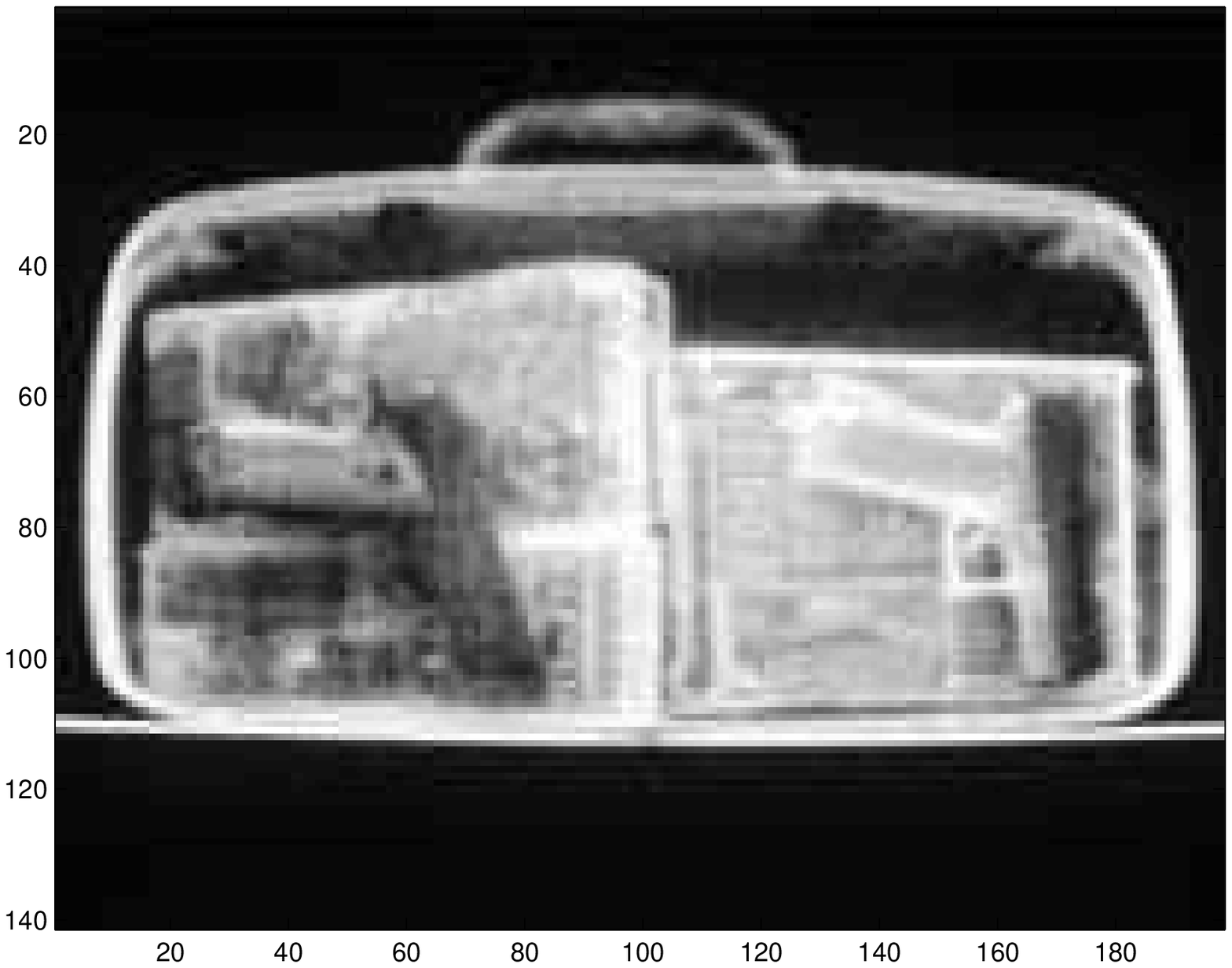,width=3.5 cm}} & 
\mbox{\epsfig{file=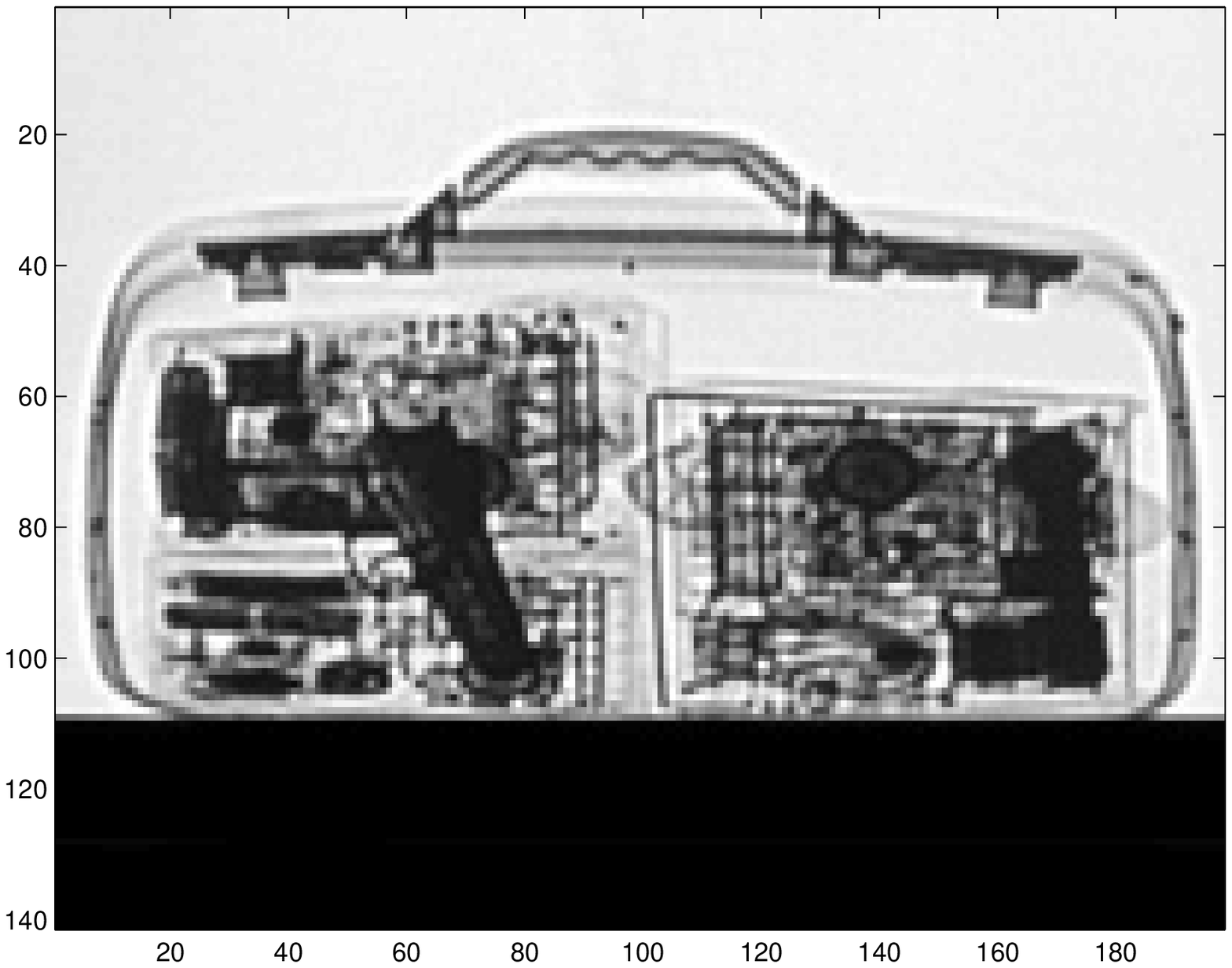,width=3.5 cm}} &
\mbox{\epsfig{file=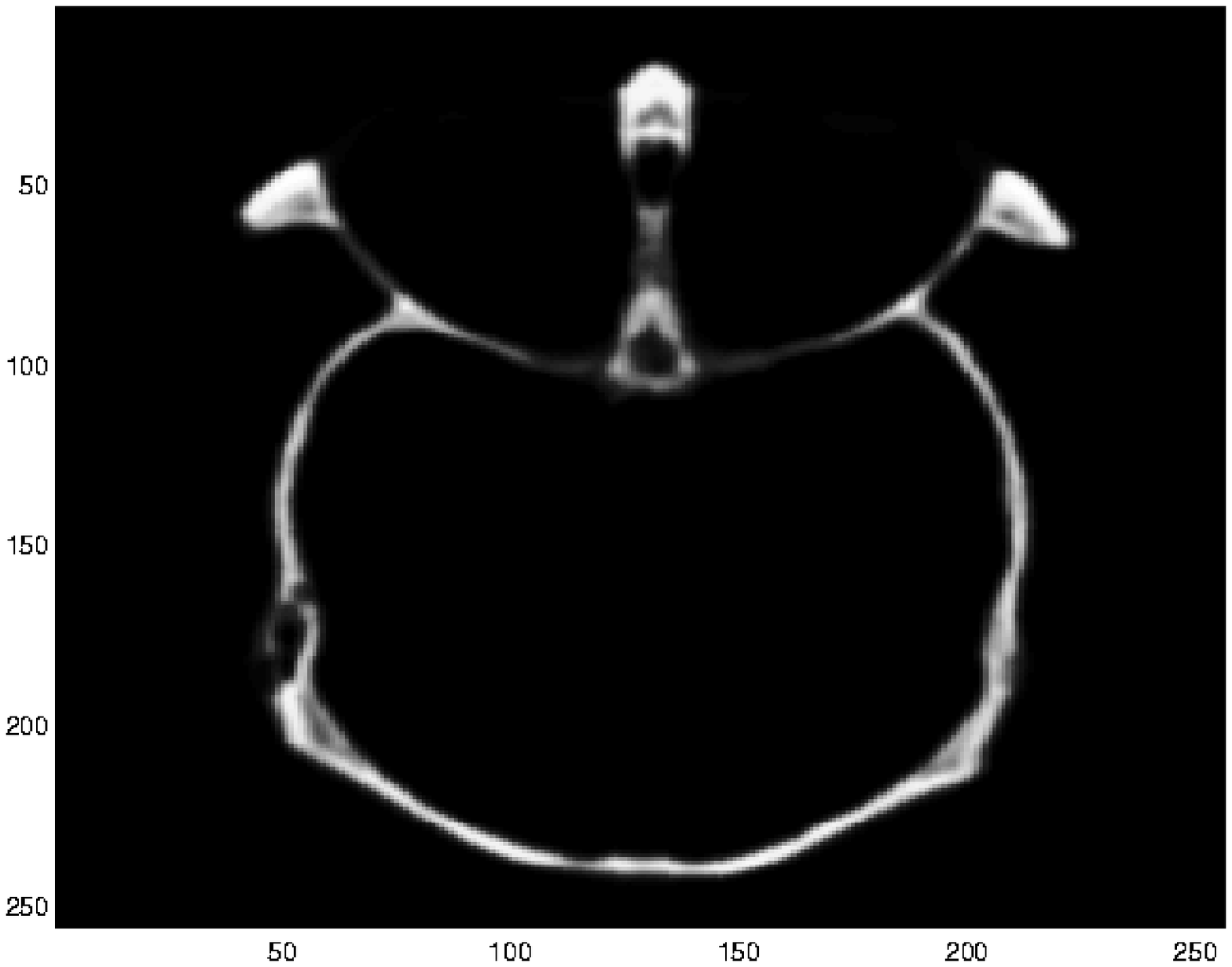,width=3.5 cm}} & 
\mbox{\epsfig{file=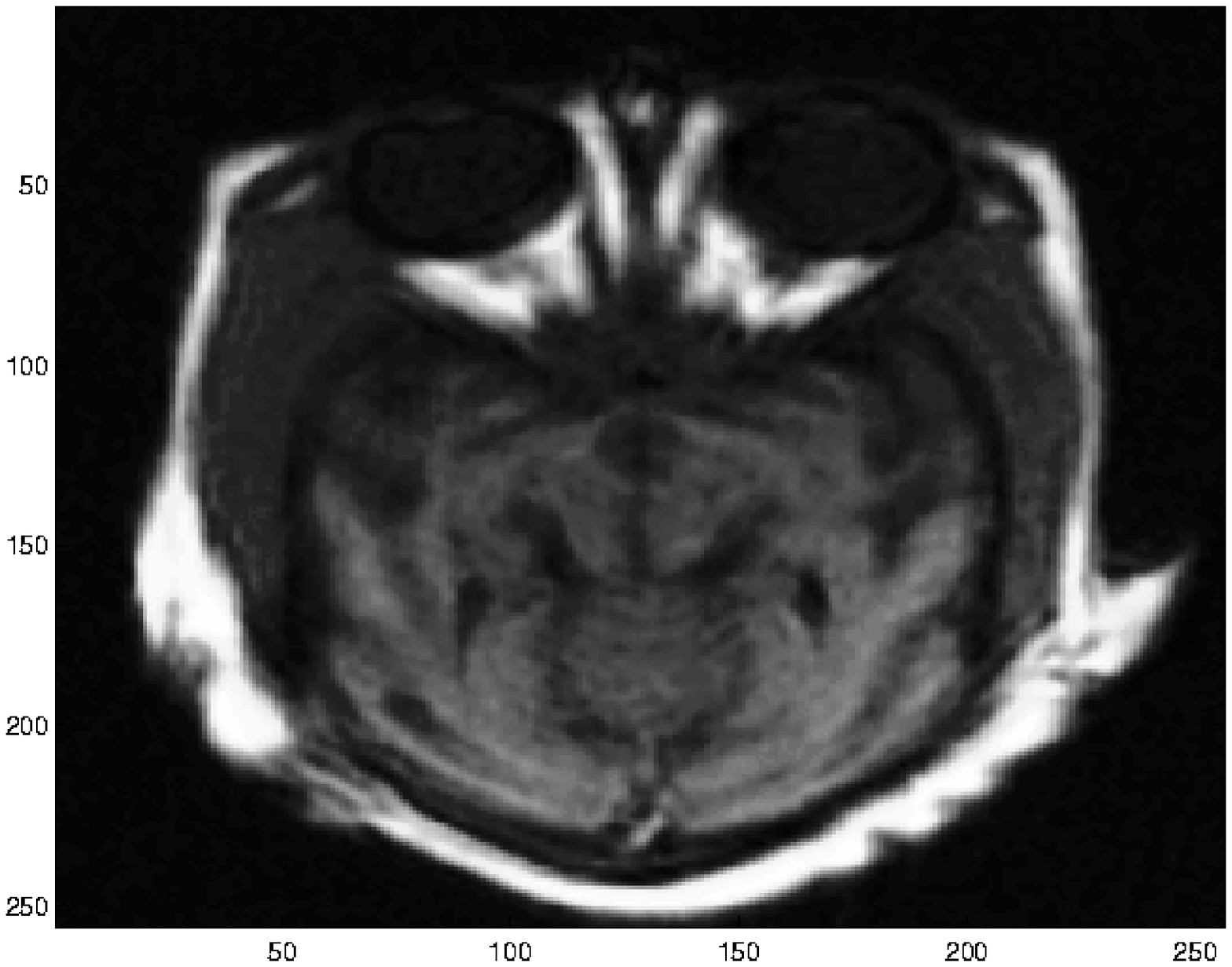,width=3.5 cm}} 
\\
 (a) & (b) & (c) & (d)
\end{tabular}

\caption{Examples of images for data fusion. a,b) Copyright American Science and Engineering, Inc., 2003 : two observations from transmission and backscattering X rays, c,d) MRI (Magnetic Resonance Imaging) and CT (Computed Tomography) images in medical imaging.}
\end{figure}

\noindent
In this paper, to show the same ideas, we consider first the case where the two measuring systems can be assumed almost perfect which means that we can write
\beq
g_i(r)=f_i(r)+\varepsilon_i(r), \qquad i=1,2
\eeq
\noindent
and we focus on defining what we mean by a common feature $z$ of the two images, how to model the relation between $f_i$ and $z$ and how to estimate $f_1$, $f_2$ and $z$ directly from the two data sets $g_1$ and $g_2$. \\
\noindent
The applications we have in mind in this work are either medical imaging or security systems imaging. As an example of the two data sets in the first application we consider MRI and CT images and as an example of the second application we consider a transmission and a backscattering diffusion images using X rays (see figure 1). \\
\noindent
The rest of the paper is organized as follows : In section 2 we introduce the common feature $z$, model the relation between the images $f_i$ to it through $p(f_i|z)$ and its proper characteristics through a prior law $p(z)$, and describe the Bayesian approach to estimate $f_1$, $f_2$ and $z$ through the \apost law $p(f_1,f_2,z|g_1,g_2)$. In section 3 we give some details on the selection of \aprio probability laws $p(\thetab)$ of the hyperparameters which define the \apost law $p(f_1,f_2,z|g_1,g_2)$. In section 4 we give detailed expressions of the aforementionned \apost law and propose general structure of the MCMC algorithm to estimate $f_1$, $f_2$ and $z$. Finally, in section 5 we present some simulation results to show the performances of the proposed method.

\section{Modeling for data fusion}

\noindent
In this paper we consider the model (2) where after discretization and using the notations $\gb_i=[g_i(1),\dots,g_i(S)]^T$, $\fb_i=[f_i(1),\dots,f_i(S)]^T$ and $\varepsilonb_i=[\varepsilon_i(1),\dots,\varepsilon_i(S)]^T$ with $S$ the total number of pixels of the images $f_i$, we have :

\beq
\gb_i =  \fb_i+\varepsilonb_i, \qquad i=1,2 
\eeq

\noindent
Within this model and assuming Gaussian independant noises,  $p(\varepsilonb_i)=\mathcal N(0,\sigma_{\varepsilon_i}^2 \Ib)$, we have

\beq
p(\gb_1,\gb_2|\fb_1,\fb_2)  = \prod_{i=1}^2 p(\gb_i|\fb_i) = \prod_{i=1}^2 p_{\varepsilon_i}(\gb_i-\fb_i)
\nonumber
\eeq

\noindent
As we want to reconstruct an image with statistically homogeneous regions, it is natural to introduce a hidden variable $\zb=(z(1),\dots,z(S)) \in \{1,\dots,K\}^S$ which represents a common classification of the two images $f_i$. The problem is now to estimate the set of variables $(\fb_1,\fb_2,\zb)$ using the Bayesian approach :

\begin{eqnarray*}
p(\fb_1,\fb_2,\zb|\gb_1,\gb_2) & = & p(\fb_1,\fb_2|\zb,\gb_1,\gb_2)p(\zb|\gb_1,\gb_2) \\
& \propto & p(\gb_1|\fb_1,\zb) p(\gb_2|\fb_2,\zb)p(\fb_1|\zb)p(\fb_2|\zb)p(\gb_1|\zb)p(\gb_2|\zb)p(\zb) \\
& \propto & p(\zb)\prod_{i=1}^2 p(\gb_i|\fb_i)p(\fb_i|\zb)p(\gb_i|\zb) 
\end{eqnarray*}

\noindent
Thus to be able to give an expression for $p(\fb_1,\fb_2,\zb|\gb_1,\gb_2)$ we need to define $p(\gb_i|\fb_i)$, $p(\fb_i|z)$, $p(\gb_i|z)$ and $p(\zb)$. \\
\noindent
Assuming $\varepsilonb_i$ centered, white and Gaussian, we have :
\begin{eqnarray*}
p(\gb_i|\fb_i) & = & \mathcal N(\fb_i,\sigma_{\varepsilonb_i}^2 \Ib) \\
& = & \left( \frac{1}{2\pi \sigma_{\varepsilonb_i}^2} \right)^{\frac{S}{2}} \exp \left\{-\frac{1}{2\sigma_{\varepsilonb_i}^2}||\gb_i-\fb_i||^2 \right\}
\end{eqnarray*}

\noindent
To assign $p(\fb_i|\zb)$ we first define the sets of pixels which are in the same class :
\begin{eqnarray*}
R_k & = & \{r : z(r) = k\}, \qquad |R_k|=n_k \\
{\fb_i}_k & = & \{ f_i(r) : z(r)=k \}
\end{eqnarray*}

\noindent
Then we assume that all the pixels of an image $f_i$ which are in the same class will be characterized by a mean ${m_i}_k$  and a variance ${\sigma_i^2}_k$ :

\beq
p(f_i(r)|z(r)=k)=\mathcal N({m_i}_k,{\sigma_i^2}_k)
\nonumber
\eeq 

\noindent
With these notations we have :

\begin{eqnarray*}
p({\fb_i}_k) & = & \mathcal N({m_i}_k \bm{1},{\sigma_i^2}_k \Ib) \\
p(\fb_i |\zb) & = & \prod_{k=1}^K \Nc({m_i}_k \bm{1},{\sigma_i^2}_k \Ib) \\
& = & \prod_{k=1}^K \left( \frac{1}{\sqrt{2\pi {\sigma_i^2}_k}}\right)^{n_k} \exp \left\{ - \frac{1}{2 {\sigma_i^2}_k} || {\fb_i}_k-{m_i}_k \bm{1} ||^2 \right\}, \qquad i=1,2.
\end{eqnarray*}

\noindent
The next step is to define $p(\gb_i|\zb)$. To do this we may use the relation (3) and the laws $p(\fb_i|\zb)$ and $p(\varepsilonb_i)$ to obtain
\begin{eqnarray*}
p(g_i(r)|z(r)=k) & = & \mathcal N({m_i}_k, {\sigma_i^2}_k + \sigma_{\varepsilon_i}^2) \\
& = & \frac{1}{\sqrt{2 \pi ({\sigma_i^2}_k + \sigma_{\varepsilon_i}^2)}} \exp \left\{-\frac{1}{2({\sigma_i^2}_k + \sigma_{\varepsilon_i}^2)} (g_i(r)-{m_i}_k)^2\right\}
\end{eqnarray*}

\noindent
Finally we have to assign $p(\zb)$. As we introduced the hidden variable $\zb$ for finding statistically homogeneous regions in images, it is natural to define a spatial dependance on these labels. The simplest model to account for this desired local spatial dependancy is a Potts Markov Random Field model :

\beq
p(\zb) = \frac{1}{T(\alpha)} \exp \left\{ \alpha \sum_{r\in \mathcal S} \sum_{s \in \mathcal V(r)}\delta (z(r) - z(s)) \right\},
\nonumber
\eeq
where $\mathcal S$ is the set of pixels, $\delta(0)=1$, $\delta(t)=0$ si $t\neq0$, $\mathcal V(r)$ denotes the neighborhood of the pixel $r$ (here we consider a neighborhood of 4 pixels) and $\alpha$ represents the degree of the spatial dependance of the variable $\zb$. This parameter wxill be studied in the next section and fixed for our algorithm.
\noindent
We have now all the necessary prior laws $p(\gb_i|\fb_i)$, $p(\fb_i|z)$, $p(\gb_i|z)$ and $p(\zb)$ and then we can give an expression for $p(\fb_1,\fb_2,\zb|\gb_1,\gb_2)$. However these probability laws have in general unknown parameters such as $\sigma^2_{\varepsilon_i}$ in $p(\gb_i|\fb_i)$ or ${m_i}_k$ and ${\sigma_i^2}_k$ in $p(\fb_i|\zb)$. In a full Bayesian approach, we have to assign prior laws to these "hyperparameters". This point is addressed in the next section.


\section{Spatial dependance parameter of the Potts model}
In the Potts Markov Random Field (PMRF) model, the parameter $\alpha$ determines the spatial dependancy between the pixels. With this model we can expect for controlling the size of the homogeneous regions in images. Indeed, if we take for example $\alpha=0$ then we consider the pixels independant, and if we increase the value of $\alpha$, we increase the spatial dependancy. In this section we study with simulations how this spatial dependancy involves with the value of $\alpha$.

\begin{figure}[h]
\centering
\begin{tabular}{ccc}
(a) : $\alpha=0.5$ & (b) $\alpha=0.6$ & (c) $\alpha=0.65$ \\
	\mbox{\epsfig{file=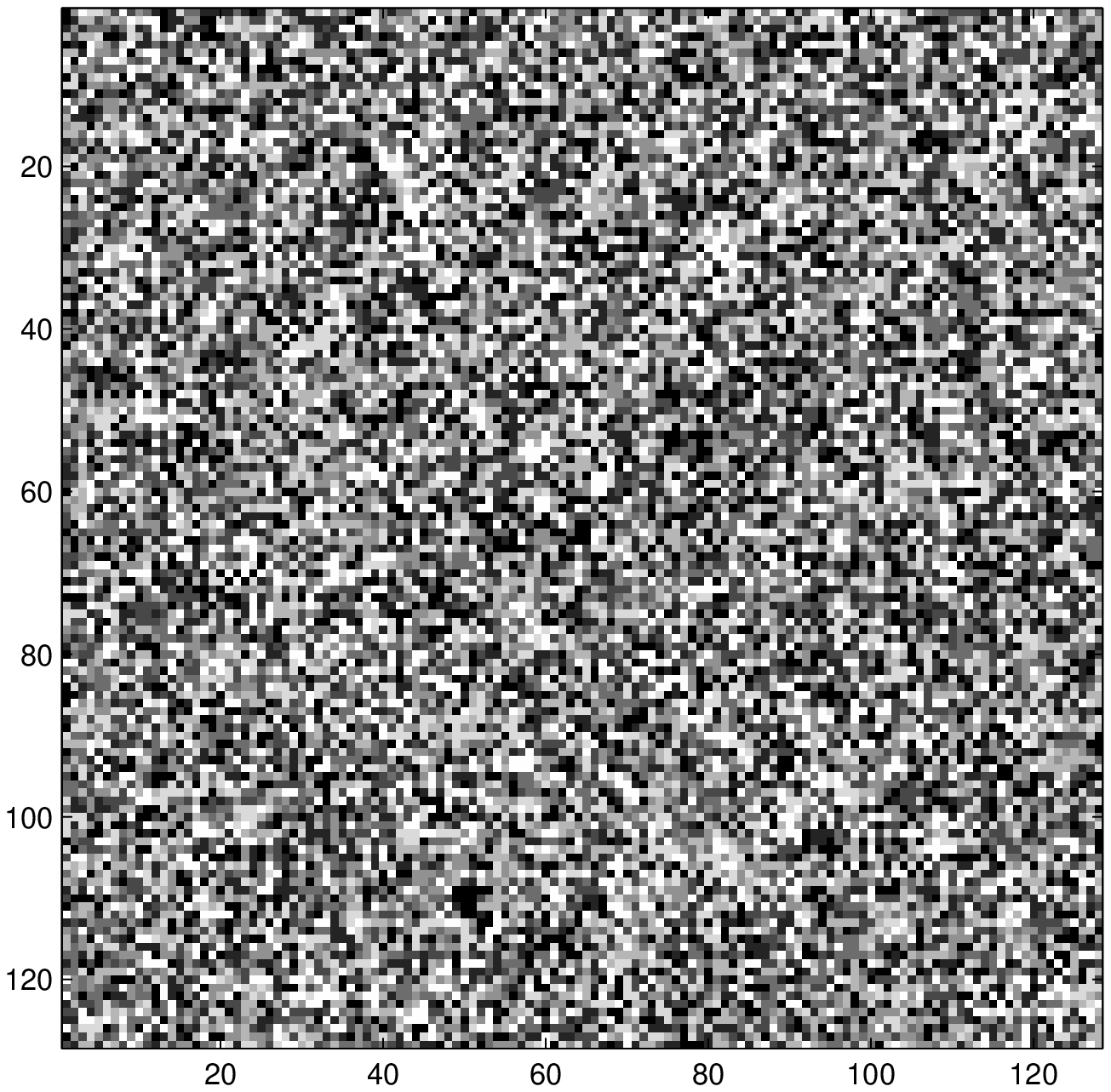,width=3.2 cm}} & 
	\mbox{\epsfig{file=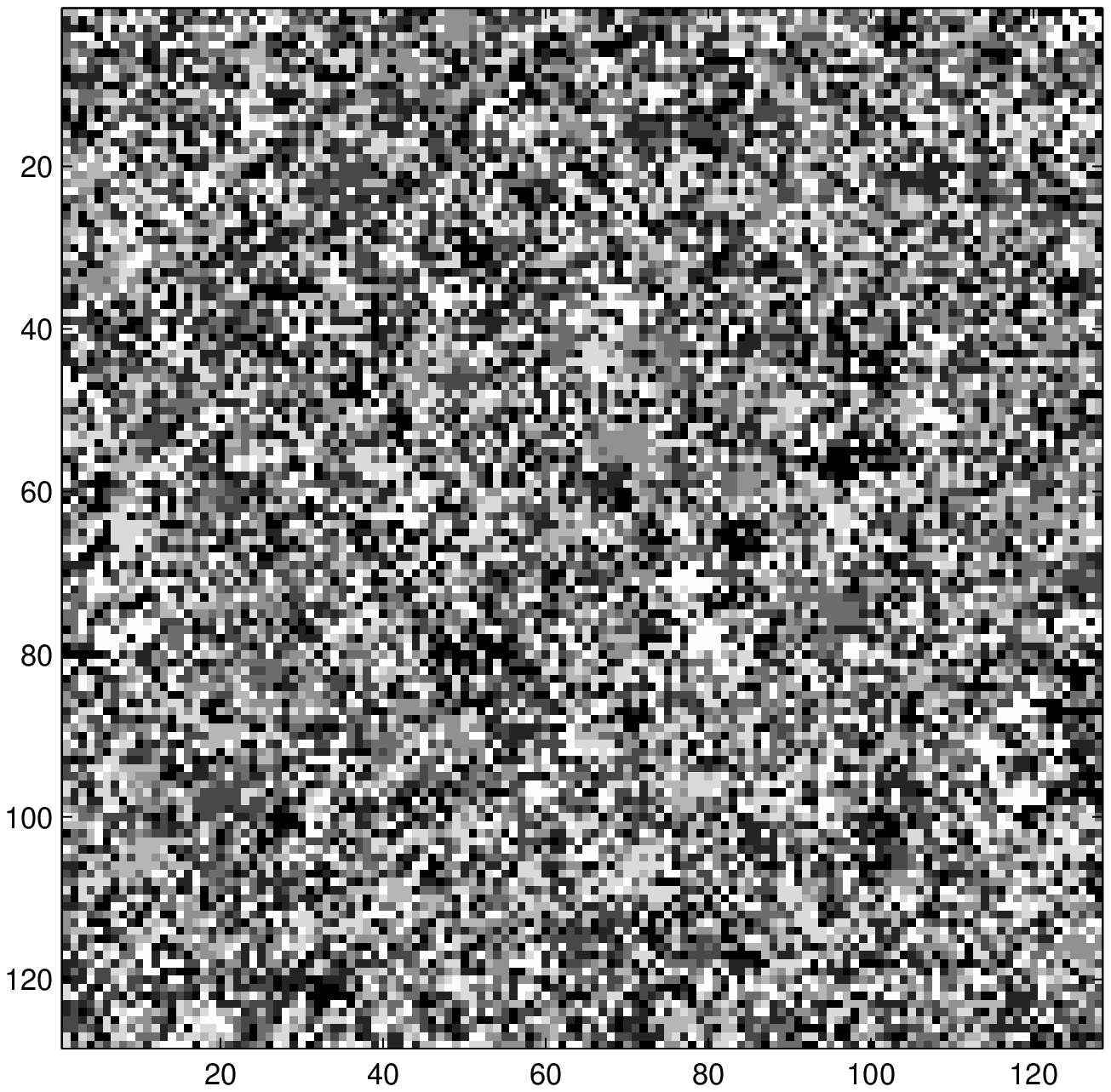,width=3.2 cm}} &
	\mbox{\epsfig{file=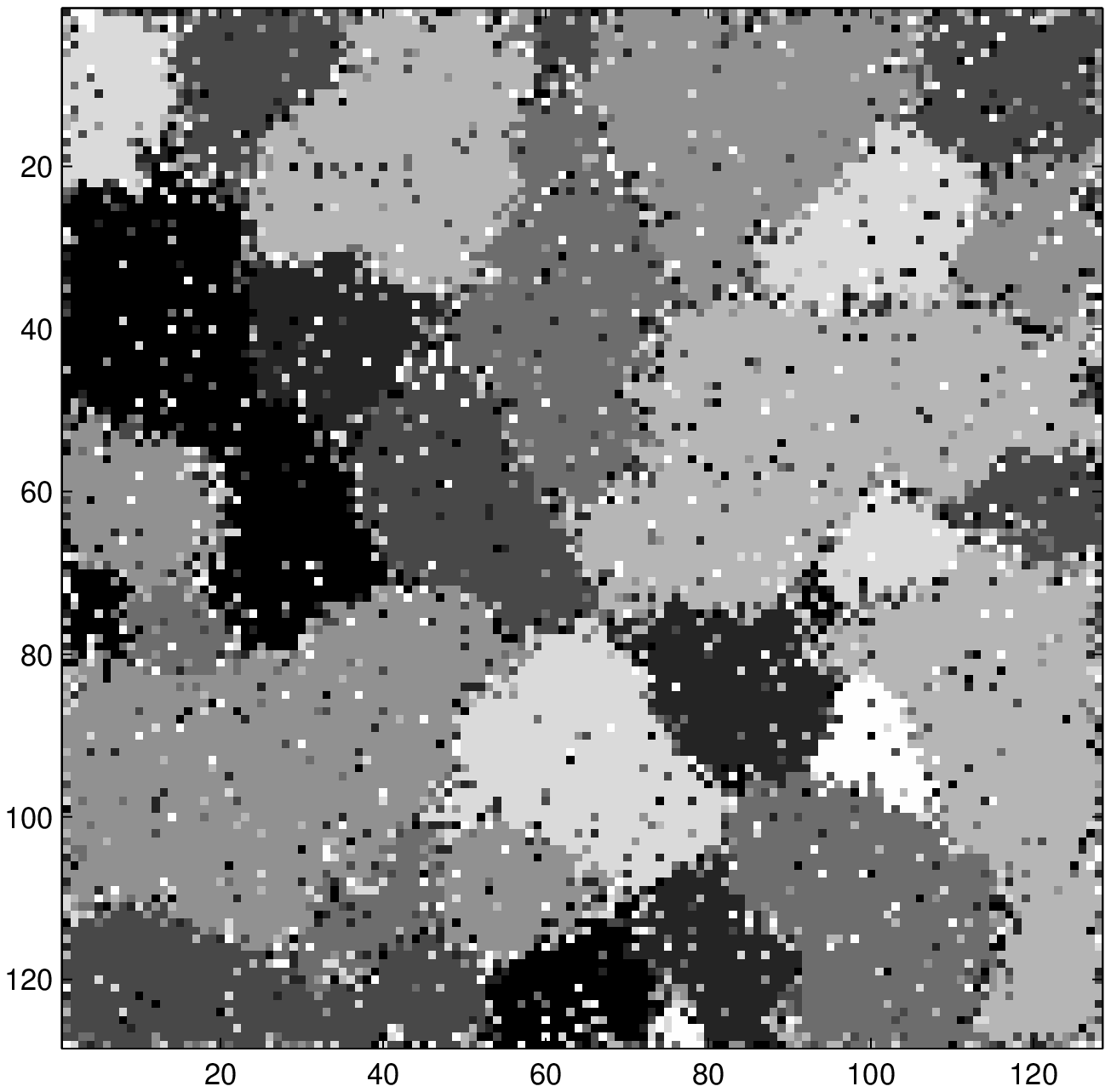,width=3.2 cm}} 
\end{tabular} 
\noindent
\\
\begin{tabular}{ccc}
(d) $\alpha=0.7$ & (e) $\alpha=1$ & (f) $\alpha=1.3$ \\
\mbox{\epsfig{file=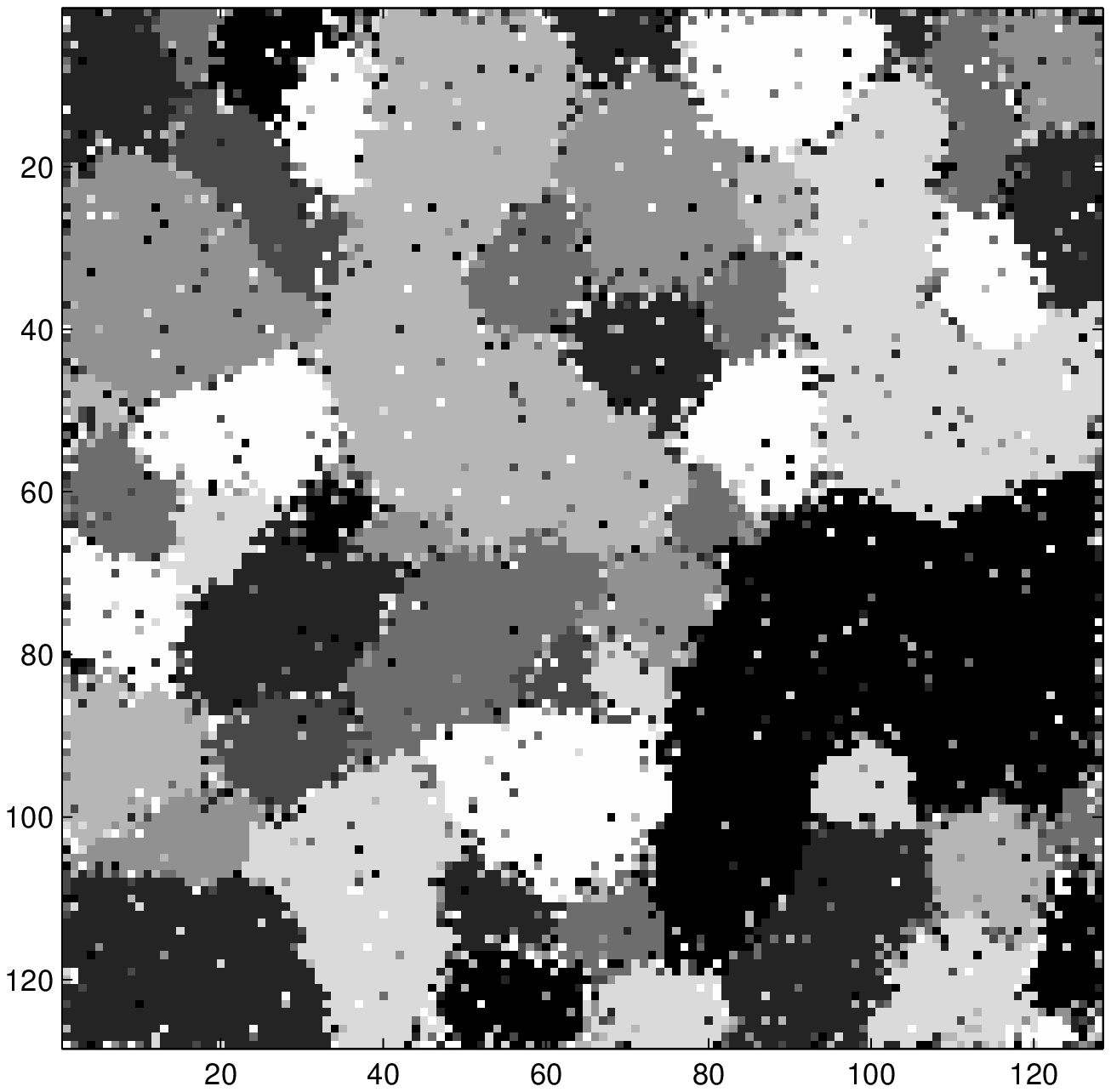,width=3.2 cm}} & 
\mbox{\epsfig{file=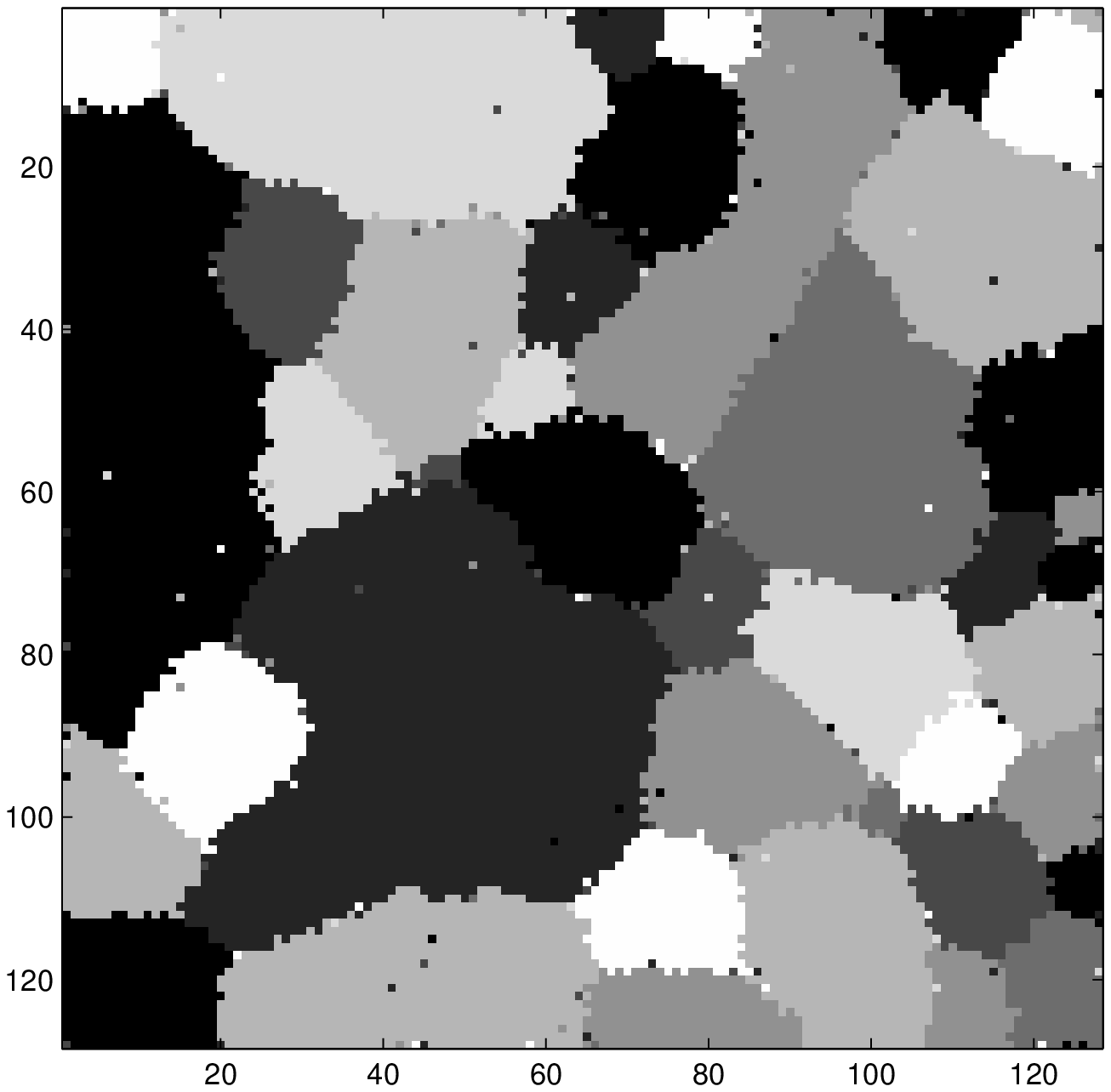,width=3.2 cm}} &	
\mbox{\epsfig{file=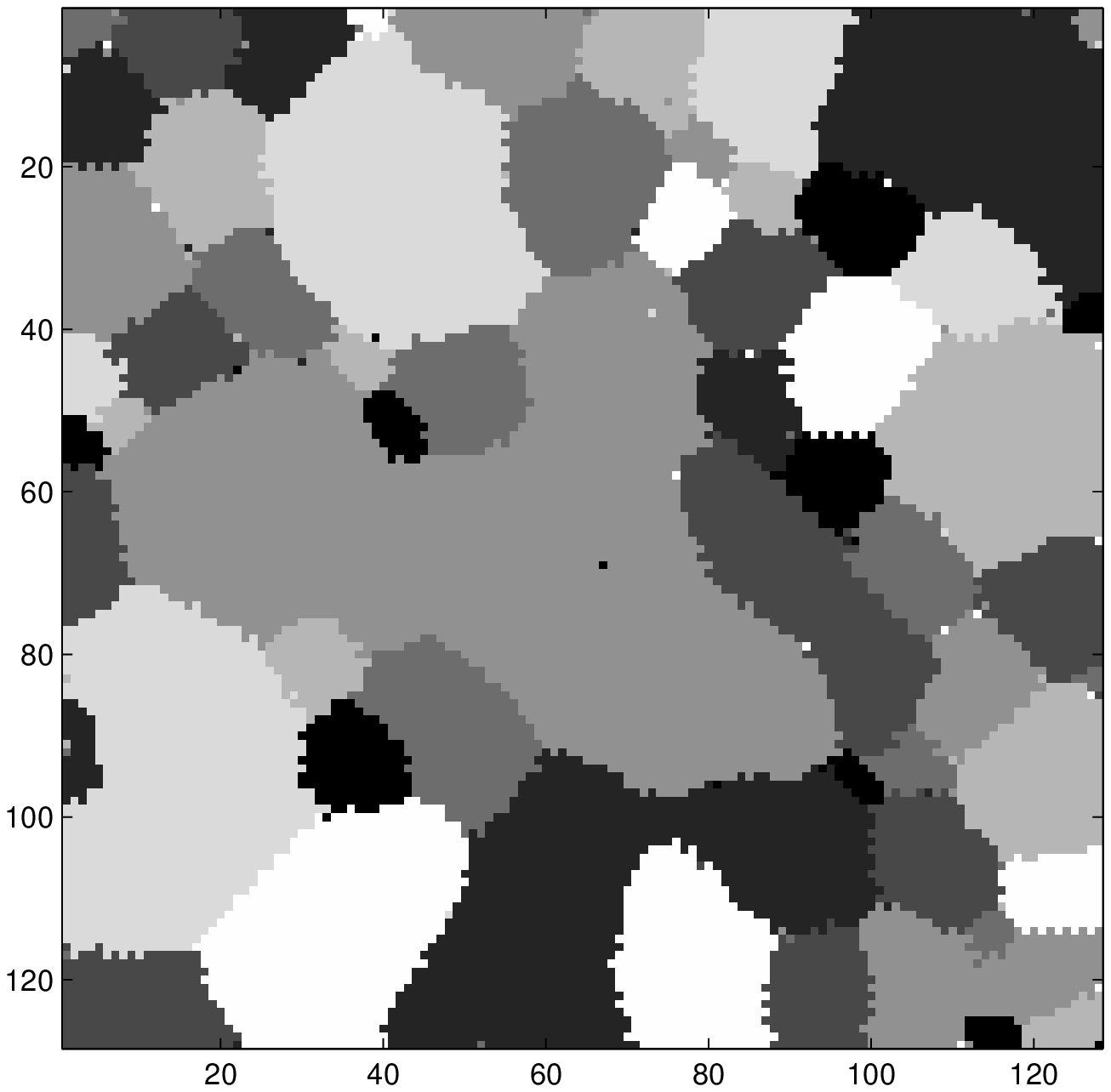,width=3.2 cm}} 
\end{tabular}
\caption{Simulations of the PRMF with variation of $\alpha$ }
\label{Potts}
\end{figure}

Figure \ref{Potts} represents some realisations of the PRMF on (128$\times$128) images of 8 labels. We can see easily that the size of the homogeneous regions does not increase linearly with the value of $\alpha$. Also the simulations show the existence of a threshold between $0.6$ and $0.7$. Under this threshold the simultations of the PMRF remain strongly disturbed. Beyond this threshold some labels become prevalent and the homogeneous regions become quickly very large. \\
The PMRF model is often used in image processing and gives a possibility to control the size of the homogeneous regions. In our algorithm we will fix the value of $\alpha$ to 1.3 for introducing a strong spatial dependancy between labels.


\section{Prior selection of the hyperparameters}
\noindent
There is an extensive litterature on the construction of non informative priors. In this section we use results from \cite{Snoussi02c} to choose particular priors, taking into account the restriction of our particular parmetrical model. In \cite{Snoussi02c} the authors used differential geometry tools to construct particular priors. \\
\noindent
Let $\mb_i=({m_i}_k)_{k=1,\dots,K}$ and $\sigmab_i^2=({\sigma_i^2}_k)_{k=1,\dots,K}$ be the means and the variances of the pixels in different regions of the images $\fb_i$ as defined before. We define $\thetab_i$ as the set of all the parameters which must be estimated :
\beq
\thetab_i = (\sigma_{\varepsilon_i}^2,\mb_i,\sigmab_i^2), \quad i=1,2
\nonumber
\eeq
We choose the prior distribution of $\thetab_i$ of the following form :
\beq
\pi(\thetab)\propto e^{-C D_\delta (p_\thetab,p_0)}\sqrt{||g(\thetab)||},
\nonumber
\eeq 
where $p_\thetab$ is the likelihood of $\thetab$, $p_0$ is a reference distribution, $C$ is a constant which represents the confidence degree we have on $p_0$, $D_\delta$ is the $\delta-divergence$ and $g$ is the Fisher information matrix. \\
The authors in (\cite{Snoussi02c}) showed that if we choose this prior distribution for $\thetab_i$ with $\delta=0$, we find the conjugate priors. When applied those results for our case, where these priors become : \\
\indent
- Inverse Gamma $\mathcal{IG}(\alpha^{\varepsilon_i}_0,\beta^{\varepsilon_i}_0)$ and $\mathcal{IG}({\alpha_i}_0,{\beta_i}_0)$ respectively for the variances $\sigma_{\varepsilon_i}^2$ and ${\sigma^2_i}_k$, \\
\indent
- Gaussian $\mathcal N({m_i}_0,{\sigma^2_i}_0)$ for the means ${m_i}_k$. \\ 
\\
\noindent
The hyper-hyperparameters ${\alpha_i}_0$, ${\beta_i}_0$, ${m_i}_0$ and ${\sigma^2_i}_0$ are fixed and the results are not in general too sensitive to their exact values.


\section{\apost distributions for the Gibbs algorithm}
\noindent
The Bayesian approach consists now to estimate the whole set of variables $(\fb_1,\fb_2,\zb,\thetab_1,\thetab_2)$ following the joint \apost distribution $p(\fb_1,\fb_2,\zb,\thetab_1,\thetab_2 | \gb_1,\gb_2)$. It is difficult to simulate a joint sample $(\hat{\fb_1},\hat{\fb_2},\hat{\zb},\hat{\thetab_1},\hat{\thetab_2})$ directly from his joint \apost distribution. However we can note that considering the prior laws defined before, we are able to simulate the conditionnal \apost laws $p(\fb_1,\fb_2,\zb| \gb_1,\gb_2,\thetab_1,\thetab_2 )$ and $p(\thetab_1,\thetab_2 | \gb_1,\gb_2,\fb_1,\fb_2,\zb)$. That's why we propose a Gibbs algorithm to estimate $(\hat{\fb_1},\hat{\fb_2},\hat{\zb},\hat{\thetab_1},\hat{\thetab_2})$, decomposing this set of variables into two subsets, $(\fb_1,\fb_2,\zb)$ and $(\thetab_1,\thetab_2)$. Then the Gibbs algorithm follows : given an initial state $(\hat{\thetab_1},\hat{\thetab_2})^{(0)}$,
\begin{center}
{\bf Gibbs sampling} \\ 
\begin{tabular}{l}
\hline   repeat until convergence \\ \\
 1. simulate $(\hat{\fb_1}^{(n)},\hat{\fb_2}^{(n)},\hat{\zb}^{(n)}) \sim p(\fb_1,\fb_2,\zb | \gb_1,\gb_2,\hat{\thetab_1}^{(n-1)},\hat{\thetab_2}^{(n-1)})$ \\
 2. simulate $\hat{\thetab_i}^{(n)} \sim p(\thetab_i|\gb_i,\hat{\fb_i}^{(n)},\hat{\zb}^{(n)})$ \\
\hline
\end{tabular}
\end{center}

\noindent 
\\ \\
\noindent
We will now define the conditionnal \apost distribution we use for the Gibbs algorithm. \\

\noindent
{\bf sampling $\thetab_i|\fb_i,\gb_i,\zb$ :} \\ \\
We have the following relation :
\beq 
p(\thetab_i|\fb_i,\gb_i,\zb) \propto p({\sigma_{\varepsilon_i}^2} | \fb_i,\gb_i)\hbox{ }p(\mb_i,\sigmab^2_i | \fb_i,\zb)
\nonumber
\eeq
\noindent 
and then we can use the Bayes formula :
\beq
p(\mb_i,\sigmab_i^2|\fb_i,\zb) \propto p(\fb_i|\zb,\mb_i,\sigmab_i^2)p(\mb_i)p(\sigmab_i^2)
\nonumber
\eeq
\noindent
and
\beq
p(\sigma_{\varepsilon_i}^2|\fb_i,\gb_i) \propto p(\gb_i|\fb_i,\sigma_{\varepsilon_i}^2)p(\sigma_{\varepsilon_i}^2)
\nonumber
\eeq
Those \apost distributions are calculated from the prior selection fixed before and we have 

\indent
- ${m_i}_k|\fb_i,\zb,{\sigma_i^2}_k,{m_i}_0, {\sigma^2_i}_0 \sim \mathcal N({\mu_i}_k,{v_i^2}_k)$, with
\begin{eqnarray*}
{\mu_i}_k & = &  {v_i^2}_k \left( \frac{{m_i}_0}{{\sigma^2_i}_0} + \frac{1}{{\sigma^2_i}_k} \sum_{r \in R_k} f_i(r)   \right) \\
{v_i^2}_k & = & \left( \frac{n_k}{{\sigma_i^2}_k} + \frac{1}{{\sigma^2_i}_0}\right)^{-1}
\end{eqnarray*} 
\indent 
- ${\sigma_i^2}_k |\fb_i,\zb,{\alpha_i}_0,{\beta_i}_0 \sim \mathcal{IG}({\alpha_i}_k,{\beta_i}_k)$, with
\begin{eqnarray*}
{\alpha_i}_k & = & {\alpha_i}_0 + \frac{n_k}{2} \\
{\beta_i}_k & = & {\beta_i}_0 + \frac{1}{2}\sum_{r \in R_k} (f_i(r)-{m_i}_k)^2
\end{eqnarray*} 
\indent
- $\sigma_{\varepsilon_i}^2 |\fb_i,\gb_i \sim \mathcal{IG}(\nu_i,\Sigma_i)$, with
\begin{eqnarray*}
\nu_i & = & \frac{S}{2} + \alpha^{\varepsilon_i}_0, \qquad S=\mbox{ total number of pixels} \\
\Sigma_i & = & \frac{1}{2} ||\gb_i-\fb_i||^2 + \beta^{\varepsilon_i}_0
\end{eqnarray*}

\noindent
{\bf Sampling $\fb_1,\fb_2,\zb|\gb_1,\gb_2,\thetab_1,\thetab_2$ :}\\ \\
Using the Bayes formula we have :
\beq 
p(\fb_1,\fb_2,\zb | \gb_1,\gb_2,\thetab_1,\thetab_2)=p(\fb_1,\fb_2 | \zb,\gb_1,\gb_2,\thetab_1,\thetab_2)p(\zb | \gb_1,\gb_2,\thetab_1,\thetab_2)
\nonumber
\eeq
\noindent
Then the sampling of this joint distribution is again obtained through a Gibbs sampling scheme and then $p(\fb_i|\gb_i,\zb,\thetab_i)$ by sampling first $p(\zb | \gb_1,\gb_2,\thetab_1,\thetab_2)$. For the first step we have :
\begin{eqnarray*}
p(\zb | \gb_1,\gb_2,\thetab_1,\thetab_2) & \propto & p(\gb_1,\gb_2 |\zb,\thetab_1,\thetab_2)p(\zb) \\
& = & p(\gb_1|\zb,\thetab_1)p(\gb_2 |\zb,\thetab_2) p(\zb)
\end{eqnarray*}
and for the next step we have :
\beq
p(f_i(r)|g_i(r),z(r)=k,\thetab_i) = \mathcal N({m_i}_k^{apost},{\sigma_i^2}_k^{apost})
\nonumber
\eeq
where
\begin{eqnarray*}
{\sigma_i^2}_k^{apost} & = & \left( \frac{1}{\sigma_{\varepsilon_i}^2} + \frac{1}{\sigma_i^2}_k\right)^{-1} \\
{m_i}_k^{apost} & = & {\sigma_i^2}_k^{apost} \left( \frac{g_i(r)}{\sigma_{\varepsilon_i}^2} + \frac{{m_i}_k}{\sigma_i^2}_k \right).
\end{eqnarray*}

\noindent
As we choosed a Potts Markov Random Field model for the labels, we may note that an exact sampling of the \apost distribution $p(\zb | \gb_1,\gb_2,\thetab_1,\thetab_2)$ is impossible. In theory, in each step, we have to implement again a third Gibbs sampling to obtain exact samples of  $\hat{\zb}$. However this will increase significantly the complexity of the algorithm. To obtain a faster algorithm, the solution we propose consists in implementing only one cycle of the Gibbs sampling for $\zb$ in each iteration. In fact it comes down to decompose the set of variables into three subsets $(\thetab_1,\thetab_2)$, $(\fb_1,\fb_2)$, and $\zb$. \\
\noindent
The Gibbs algorithm we propose is then : given an initial state $(\hat{\thetab_1},\hat{\thetab_2},\hat{\zb})^{(0)}$, 
\begin{center}
{\bf Gibbs sampling} \\
\begin{tabular}{l}
\hline  repeat until convergence \\
 \begin{tabular}{l} 
 		1. simulate $\hat{\zb}^{(n)} \sim p(\zb|\hat{\zb}^{(n-1)},\gb_1,\gb_2,\hat{\thetab_1}^{(n-1)},\hat{\thetab_2}^{(n-1)})$ \\
 		$\quad$ simulate $\hat{\fb_i}^{(n)} \sim p(\fb_i | \gb_i,\hat{\zb}^{(n)},\hat{\thetab_i}^{(n-1)})$ 
	\end{tabular} \\
 \begin{tabular}{l}
  2. simulate $\hat{\thetab_i}^{(n)} \sim p(\thetab_i|\hat{\fb_i}^{(n)},\hat{\zb}^{(n)},\gb_i)$ 
 	\end{tabular} \\
\hline
\end{tabular}
\end{center} 

\noindent
\\ \\
As we choosed a first order neighborhood system for the labels, we may also note that it is pssible to implement the Gibbs algorithm in parallel. Indeed, we can decompose the whole set of pixels into two subsets forming a chessboard (see figure 2). In this case if we fix the black (respectively white) labels, then the white (respectively black) labels become independant.
\begin{figure}[h]
\centering
\mbox{\epsfig{file=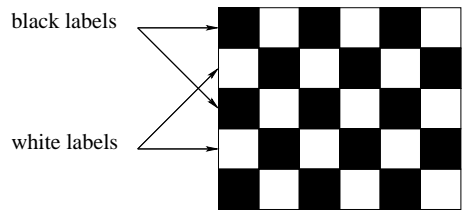,width=6 cm}}
\caption{Chessboard decomposition of the labels $z$}
\end{figure}

\noindent
This decomposition reduces the complexity of the Gibbs algorithm because we can simulate the whole set of labels in only two steps. The Parallel Gibbs algorithm we implemented is then the following : given an initial state $(\hat{\thetab_1},\hat{\thetab_2},\hat{\zb})^{(0)}$,
\newpage 
\begin{center}
{\bf  Parallel Gibbs sampling} \\
\begin{tabular}{l}
\hline repeat until convergence \\
 \begin{tabular}{l} 
 		1. simulate $\hat{\zb_B}^{(n)} \sim p(\zb|\hat{\zb_W}^{(n-1)},\gb_1,\gb_2,\hat{\thetab_1}^{(n-1)},\hat{\thetab_2}^{(n-1)})$ \\
		$\quad$ simulate $\hat{\zb_W}^{(n)} \sim p(\zb|\hat{\zb_B}^{(n)},\gb_1,\gb_2,\hat{\thetab_1}^{(n-1)},\hat{\thetab_2}^{(n-1)})$ \\
 		$\quad$ simulate $\hat{\fb_i}^{(n)} \sim p(\fb_i | \gb_i,\hat{\zb}^{(n)},\hat{\thetab_i}^{(n-1)})$ 
	\end{tabular} \\
 \begin{tabular}{l}
  2. simulate $\hat{\thetab_i}^{(n)} \sim p(\thetab_i|\hat{\fb_i}^{(n)},\hat{\zb}^{(n)},\gb_i)$ 
 	\end{tabular} \\
\hline
\end{tabular}
\end{center} 
\noindent
In the following we have implemented this algorithm.


\section{Spatial dependance in the estimated images}
\noindent
We want now to introduce a dependance between pixels of $f_i$ which are in a same homogeneous region. Our aim is to improve the reconstructed images and then (because our algorithm is iterative) improve the quality of our classification. We will now describe this new modelisation and the modifications it implies.

\subsection{New modelisation on the images $f_i$}
\noindent
We now consider that pixels $f_i(r)$ inside a same region are dependant. However pixels being in different regions stay independant. We then introduce a "contour" variable $\qb$ as follows :
\begin{eqnarray*}
q(r) & = & 0 \mbox{ if } \{z(s),s\sim r\} \mbox{ are in a same region} \\
& = & 1 \mbox{ else}
\end{eqnarray*}
Then we have the following :
\beq
p(\fb_i |\zb,\qb,\thetab_i)=\prod_{k=1}^K p({\fb_i}_k |\zb,\qb,\thetab_i)
\nonumber
\eeq
\noindent
Let note ${\fb_i}_N(r)=\{ f_i(s),s\sim r\}$. Then we can write :
\begin{eqnarray*}
p(f_i(r)|z(r)=k,q(r),{\fb_i}_N(r),\thetab_i) & = & \Nc(\mu_k,\sigma_k^2) \mbox{ if } q(r)=1 \\
& = & \Nc(\frac{1}{4} \bm{1}^t {\fb_i}_N(r),\frac{\sigma_k^2}{4})\mbox{ if } q(r)=0
\end{eqnarray*}
\noindent
Note also 
\begin{eqnarray*}
m_{f_i(r)} & = & q(r)\mu_k + (1-q(r))\frac{1}{4}\bm{1}^t {\fb_i}_N(r) \\
\sigma_{f_i(r)}^2 & = & q(r)\sigma_k^2 + (1-q(r))\frac{\sigma_k^2}{4}
\end{eqnarray*}
\noindent
Then we can write the distribution $p(g_i(r)|z(r)=k,q(r),{\fb_i}_N(r),\thetab_i)$ :
\beq
p(g_i(r)|z(r)=k,q(r),{\fb_i}_N(r),\thetab_i)=\Nc(m_{f_i(r)},\sigma_{f_i(r)}^2+\sigma_{\varepsilon_i}^2)
\nonumber
\eeq


\subsection{\apost distributions}

\subsubsection{Sampling $\fb_i | \gb_i,\zb,\qb,\thetab_i$}
\noindent
With the same method of section 5, we obtain the \apost distribution :
\beq
p(f_i(r)|g_i(r),z(r),q(r),{\fb_i}_N(r),\thetab_i)=\Nc(m_{apost},\sigma_{apost}^2),
\nonumber
\eeq
\noindent
with
\begin{eqnarray*}
m_{apost} & = & \sigma_{apost}^2 \left( \frac{g_i(r)}{\sigma_{\varepsilon_i}^2}+\frac{m_{f_i(r)}}{\sigma_{f_i(r)}^2} \right) \\
\sigma_{apost}^2 & = & \left( \frac{1}{\sigma_{\varepsilon_i}^2}+\frac{1}{\sigma_{f_i(r)}^2}\right)^{-1}
\end{eqnarray*}
\noindent
As we choose a spatial dependance between pixels $f_i(r)$, we have the same problem as for the labels. Indeed an exact sampling of the \apost law $p(\fb_i | \gb_i,\zb,\qb,\thetab_i)$ becomes impossible. The solution is, as for the labels, to decompose the set of pixels into a chessboard. Let then note ${\fb_i}_W$ and ${\fb_i}_B$ rspectively the white and black pixels $f_i(r)$. Then if we fix ${\fb_i}_W$, the pixels ${\fb_i}_B(r)$ are independant and we have
\beq
p({\fb_i}_B |\gb_i,\zb,\qb,\thetab_i) = \prod_{\mbox{\scriptsize r black}} p(f_i(r)|g_i(r),z(r),q(r),\thetab_i)
\nonumber
\eeq
and the symetric relation for ${\fb_i}_W$. \\
\noindent
This solution consists then in introducing a Gibbs algortihm for sampling $\fb_i$ and, as for the labels, we implement only one cycle of the Gibbs sampling for limiting the complexity of our algorithm.

\subsubsection{Sampling $\zb | \gb_1,\gb_2,\fb_1,\fb_2,\qb,\thetab_1,\thetab_2$}
As for the first case we use the \apost distribution :
\beq
p(\zb|\gb_1,\gb_2,\fb_1,\fb_2,\qb,\thetab_1,\thetab_2) \propto p(\gb_1|\zb,\fb_1,\qb,\thetab_1)p(\gb_2|\zb,\fb_2,\qb,\thetab_2)p(\zb)
\nonumber
\eeq
\noindent
The exact sampling of this distribution is still impossible but we can use the decomposition into a chessboard to obtain :
\begin{eqnarray*}
p(\zb_B|\zb_W,{\gb_N}_1,{\gb_N}_2,{\fb_B}_1,{\fb_B}_2,\qb_N,\thetab_1,\thetab_2) & \propto & p(\zb_B|\zb_W) \prod_{i=1}^2 p({\gb_N}_i|\zb,{\fb_i}_W,\qb_N,\thetab_i) \\
& = & p(\zb_B|\zb_W) \prod_{i=1}^2 \prod_{\mbox{\scriptsize r black}} p(g_i(r)|z(r),{\fb_N}_i(r),q(r),\thetab_i)
\end{eqnarray*}
Then we implement for this part one cycle of a Gibbs sampling, as in the first method described in section 5.

\subsubsection{Sampling $\thetab_i| \zb,\gb_i,\fb_i,\qb$}
We still use the same method to obtain the \apost distributions of the parameters of $\thetab_i$. However we have here to decompose the set $R_k$ into to subsets as follows :
\beq
R_k=R_k^0 \cup R_k^1
\nonumber
\eeq
\noindent
with $R_k^i=\{r;z(r)=k,q(r)=i\}$. Let also note $n_k^i=|R_k^i|$. With this decomposition we can calculate the \apost distributions of $\thetab_i$ :

- ${m_i}_k|\fb_i,\zb,\qb,{\sigma_i^2}_k,{m_i}_0, {\sigma^2_i}_0 \sim \mathcal N({\mu_i}_k,{v_i^2}_k)$, with
\begin{eqnarray*}
{\mu_i}_k & = &  {v_i^2}_k \left( \frac{{m_i}_0}{{\sigma^2_i}_0} + \frac{1}{{\sigma^2_i}_k} \sum_{r \in R_k^1} f_i(r)   \right) \\
{v_i^2}_k & = & \left( \frac{n_k^1}{{\sigma_i^2}_k} + \frac{1}{{\sigma^2_i}_0}\right)^{-1}
\end{eqnarray*} 
\indent 
- ${\sigma_i^2}_k |\fb_i,\zb,\qb,{\alpha_i}_0,{\beta_i}_0 \sim \mathcal{IG}({\alpha_i}_k,{\beta_i}_k)$, with
\begin{eqnarray*}
{\alpha_i}_k & = & {\alpha_i}_0 + \frac{n_k}{2} \\
{\beta_i}_k & = & {\beta_i}_0 + \frac{1}{2}\sum_{r \in R_k^1} (f_i(r)-{m_i}_k)^2 +2 \sum_{r \in R_k^0} (f_i(r)-\bm{1}^t {\fb_N}_i(r))^2
\end{eqnarray*} 

\indent
- $\sigma_{\varepsilon_i}^2 |\fb_i,\gb_i \sim \mathcal{IG}(\nu_i,\Sigma_i)$, with
\begin{eqnarray*}
\nu_i & = & \frac{S}{2} + \alpha^{\varepsilon_i}_0, \qquad S=\mbox{ total number of pixels} \\
\Sigma_i & = & \frac{1}{2} ||\gb_i-\fb_i||^2 + \beta^{\varepsilon_i}_0
\end{eqnarray*}


\subsection{New Gibbs algorithm}
The only difference between the algorithm of section 5 is the introduction of the new variable $\qb$ and in the sampling of $\fb_i$. Indeed we have here to decompose $\fb_i$ into two subsets ${\fb_i}_B$ and ${\fb_i}_W$. The Gibbs algorithm we have implemented is then :
\begin{center}
{\bf  Parallel Gibbs sampling} \\
\begin{tabular}{l}
\hline repeat until convergence \\
 \begin{tabular}{l} 
 		1. simulate $\hat{\zb_B}^{(n)} \sim p(\zb_B|\hat{\zb_W}^{(n-1)},\hat{\fb_i}_W^{(n-1)},\hat{\qb}^{(n-1)},\gb_1,\gb_2,\hat{\thetab_1}^{(n-1)},\hat{\thetab_2}^{(n-1)})$ \\
		$\quad$ simulate $\hat{\zb_W}^{(n)} \sim p(\zb_W|\hat{\zb_B}^{(n)},\hat{\fb_i}_B^{(n-1)},\hat{\qb}^{(n-1)},\gb_1,\gb_2,\hat{\thetab_1}^{(n-1)},\hat{\thetab_2}^{(n-1)})$ \\
 \end{tabular} \\
 \begin{tabular}{l}
		2. simulate $\hat{\qb}^{(n)} \sim p(\qb|\hat{\zb}^{(n)})$ \\
 \end{tabular} \\
 \begin{tabular}{l}
 		3. simulate $\hat{\fb_i}_B^{(n)} \sim p({\fb_i}_B |{\fb_i}_W^{(n-1)} \gb_i,\hat{\zb}^{(n)},\hat{\qb}^{(n)},\hat{\thetab_i}^{(n-1)})$ \\
		$\quad$ simulate $\hat{\fb_i}_W^{(n)} \sim p({\fb_i}_W |{\fb_i}_B^{(n-1)} \gb_i,\hat{\zb}^{(n)},\hat{\qb}^{(n)},\hat{\thetab_i}^{(n-1)})$ \\ 
 \end{tabular} \\
 \begin{tabular}{l}
  4. simulate $\hat{\thetab_i}^{(n)} \sim p(\thetab_i|\hat{\fb_i}^{(n)},\hat{\zb}^{(n)},\gb_i)$ 
 \end{tabular} \\
\hline
\end{tabular}
\end{center}


\section{Simulation and results}

\noindent
Here we illustrate two applications of the proposed method in cases of medical imaging and security systems. The first application is MRI and CT images of a brain which are (256 X 256) images.


\begin{figure}[h]
\centering
\begin{tabular}{cccc}
	\mbox{\epsfig{file=figures/medA_initiale.eps,width=3.5 cm}} & 
\mbox{\epsfig{file=figures/medB_initiale.eps,width=3.5 cm}} &
\mbox{\epsfig{file=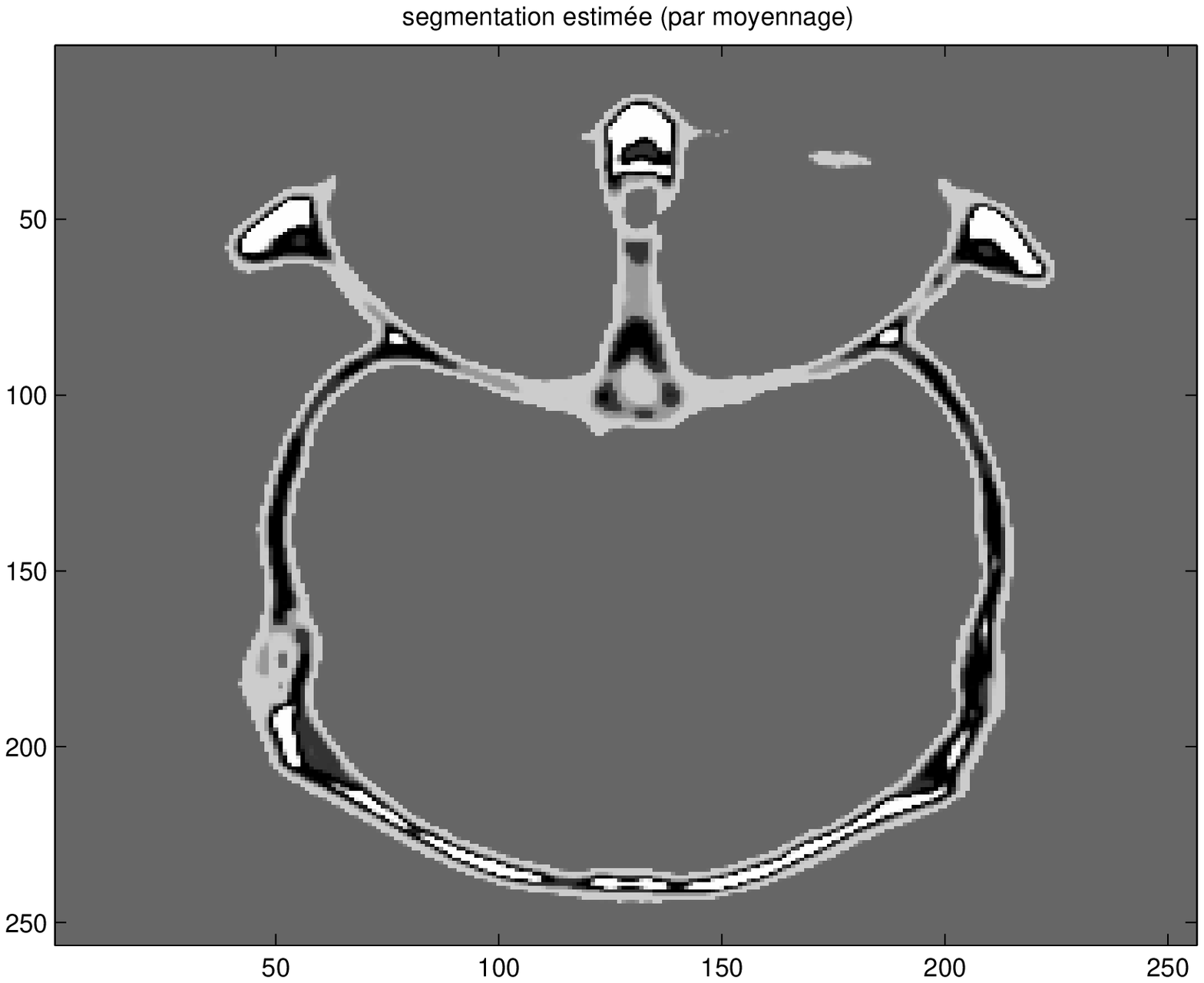,width=3.5 cm}} &
\mbox{\epsfig{file=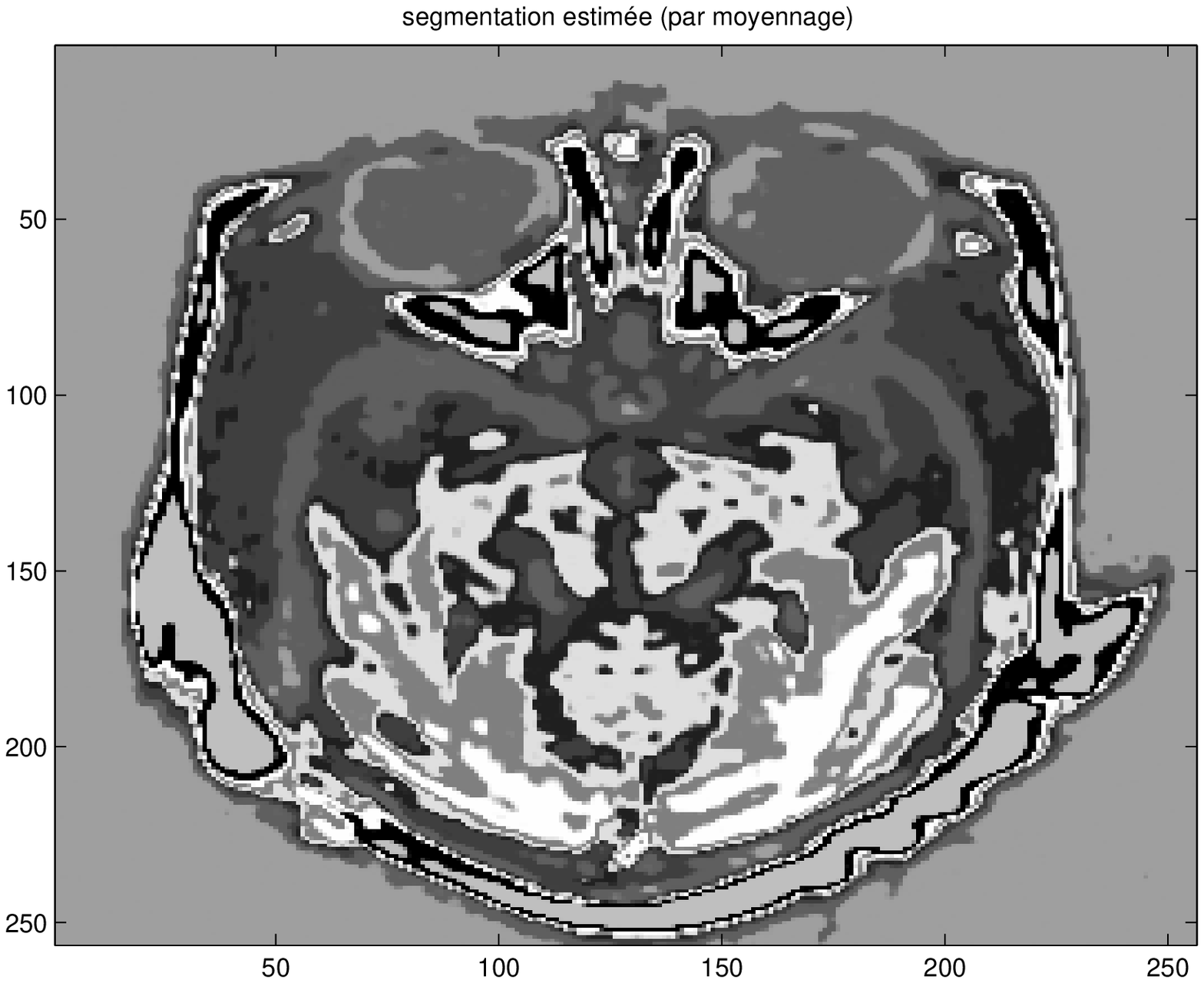,width=3.5 cm}} 

\\
 (a) & (b) & (c) & (d)
\end{tabular}

\begin{tabular}{ccc}
\mbox{\epsfig{file=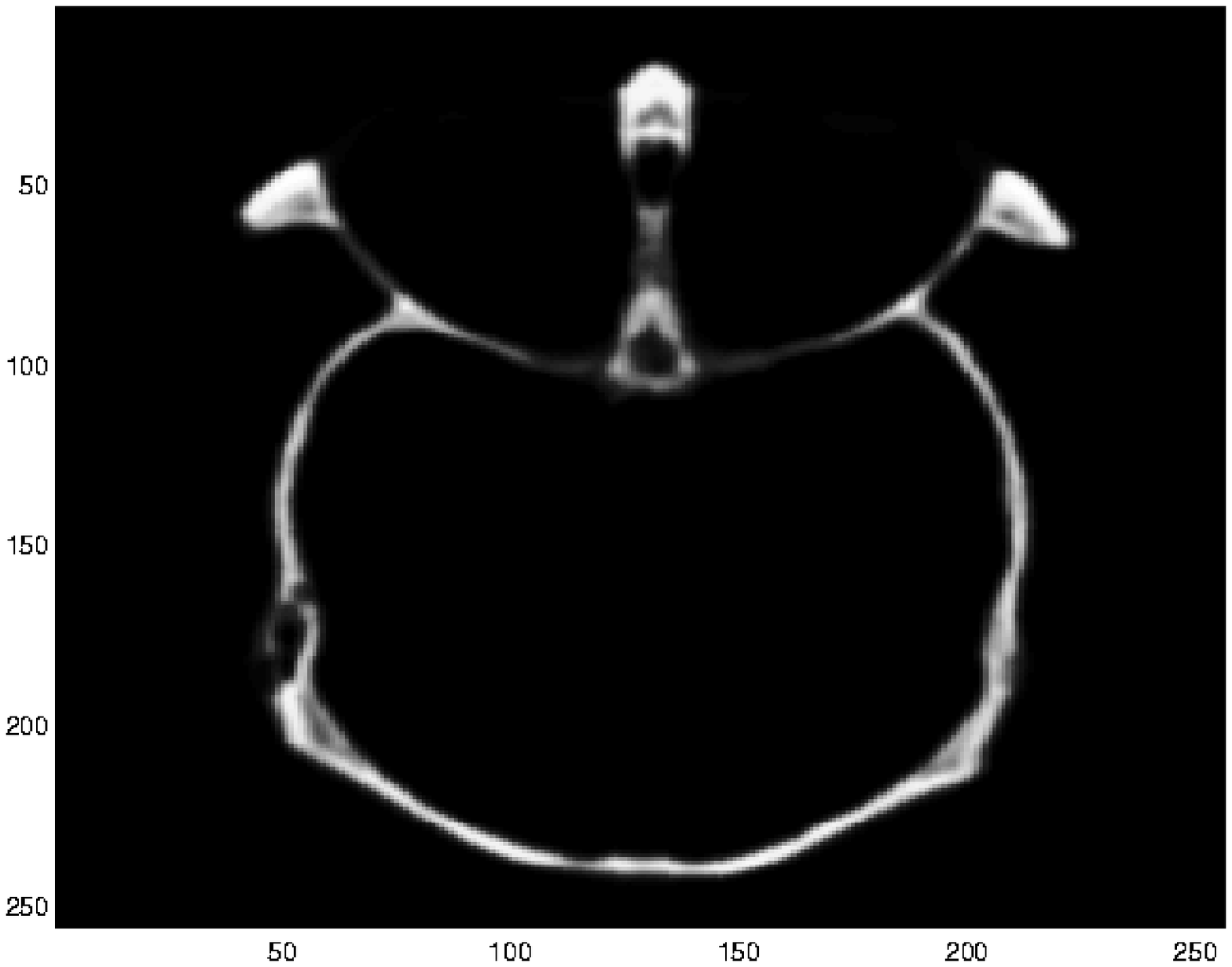,width=3.5 cm}} & 
\mbox{\epsfig{file=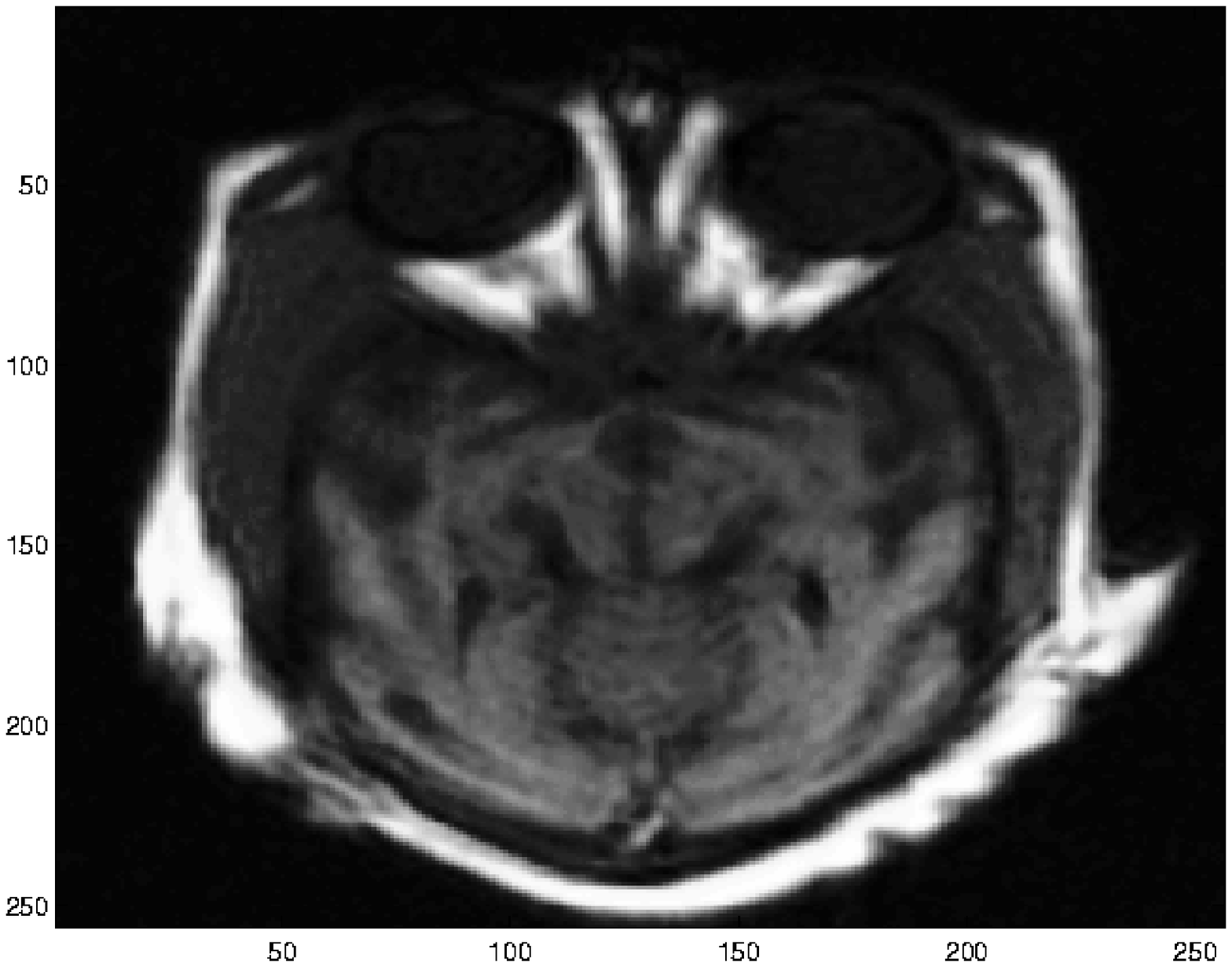,width=3.5 cm}} &	
\mbox{\epsfig{file=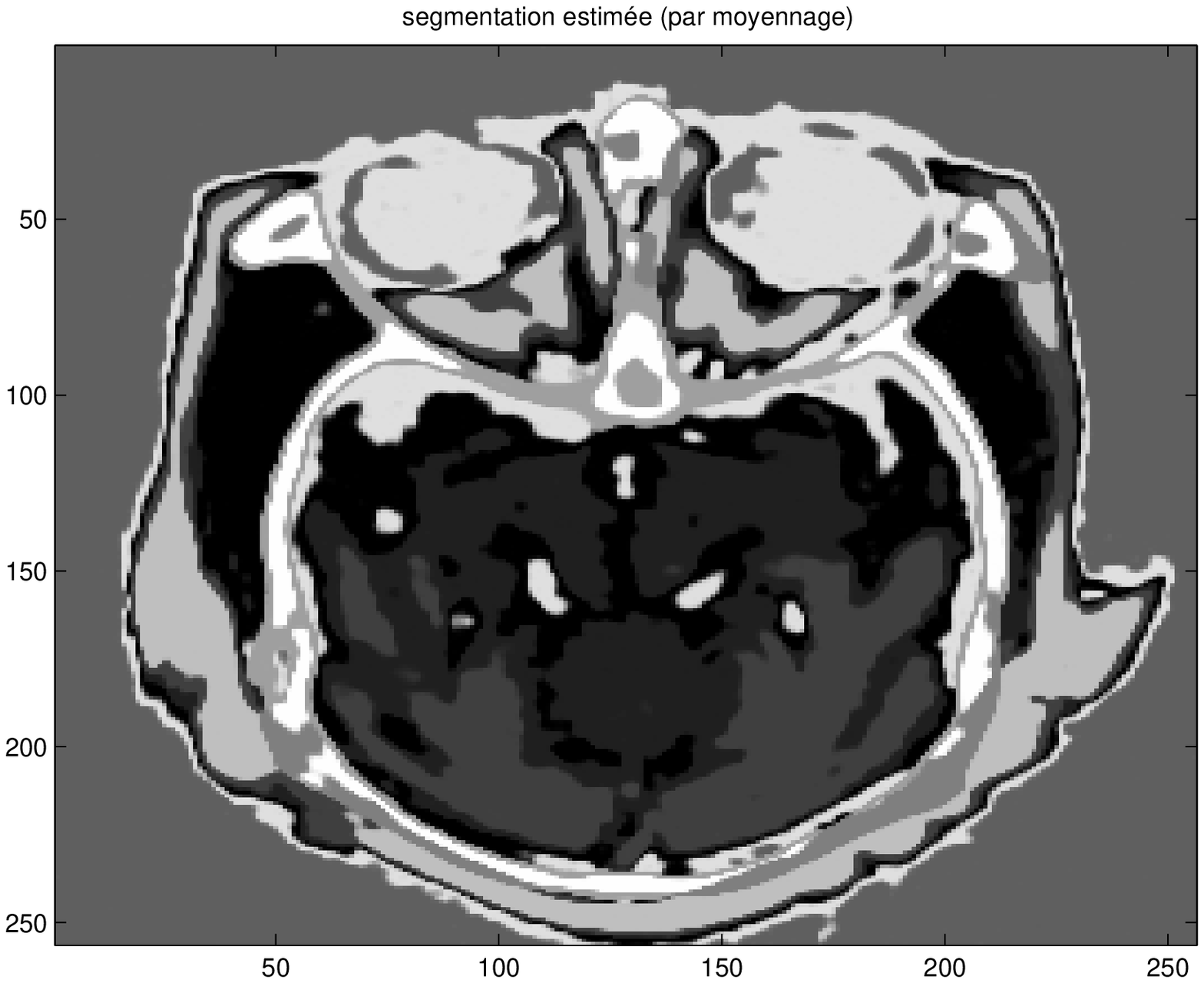,width=3.5 cm}} \\
	(e) & (f) & (g)
\end{tabular}
\caption{Results of data fusion from MRI and CT images. a,b) MRI and CT images in medical imaging. c,d) segmentation of the two images taken independantly with respectively 6 and 9 labels. e,f) respective reconstruction of the images. g) result of data fusion with 9 labels.}
\label{medical}
\end{figure}

\noindent
Figure \ref{medical} shows the data fusion result of the proposed method comparing independant segmentation of the two images and segmentation using data fusion. As it is seen on this figure the fusionned segmentation we obtain contains all the regions and boundaries of both images. This is particularly visible in the up-center of the image. \\
\noindent
The second application is X ray transmission and backscattering images, which are (141 X 192) images. The observed object is a suitcase containing two guns. Figure \ref{valise} shows the result of the proposed method. The independant segmentation of the X ray backscttering image show clearly the presence of the right gun, but it is difficult to distinguish the left gun whatever the number of labels. In this figure we can see that the X ray transmission image brings essential information to precisely distinguish the left gun without eliminating the detecton of the one on the right.


\begin{figure}
\centering
\begin{tabular}{cccc}
	\mbox{\epsfig{file=figures/valise1.eps,width=3.5 cm}} & 
\mbox{\epsfig{file=figures/valise2.eps,width=3.5 cm}} &
\mbox{\epsfig{file=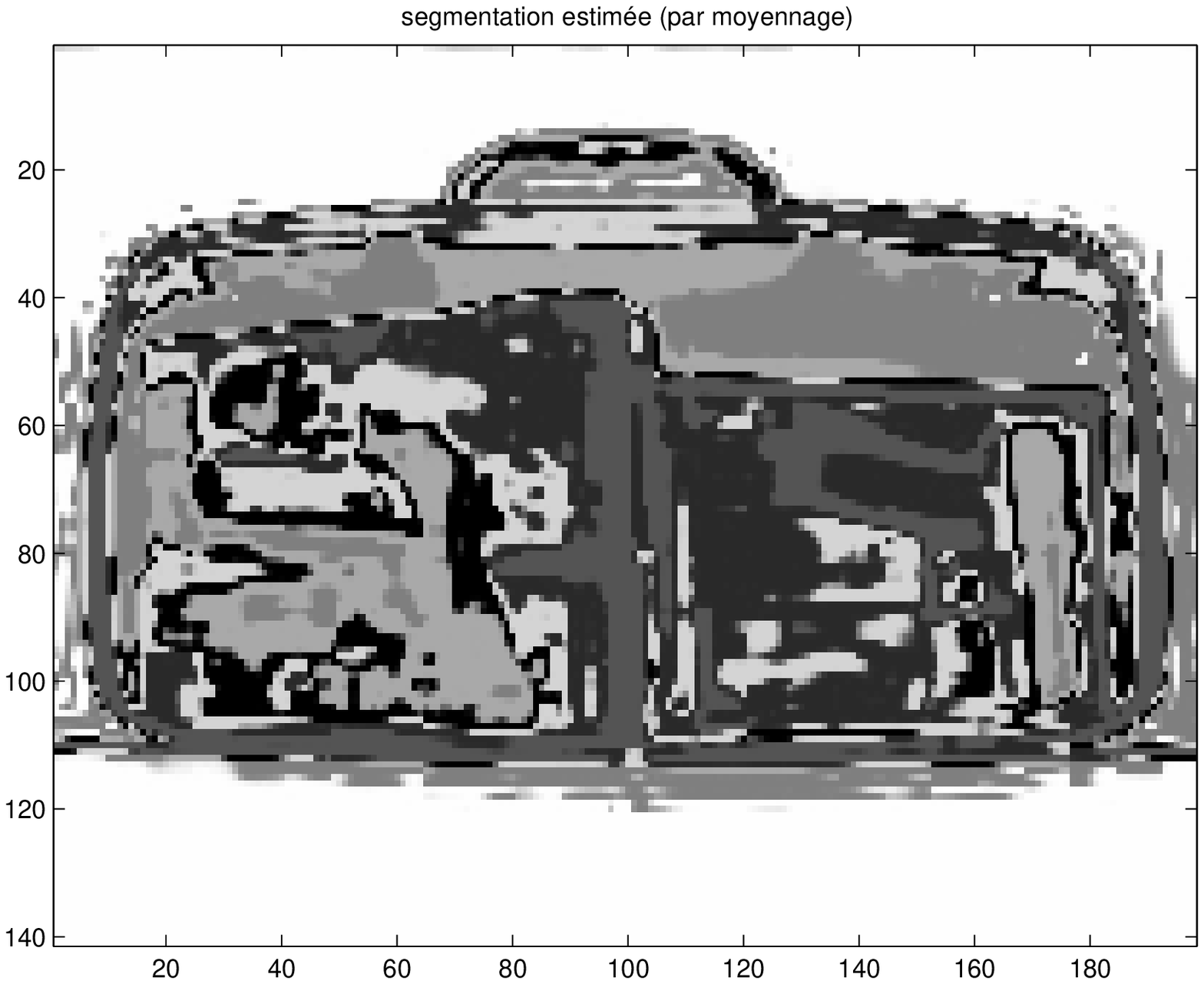,width=3.5 cm}} &
\mbox{\epsfig{file=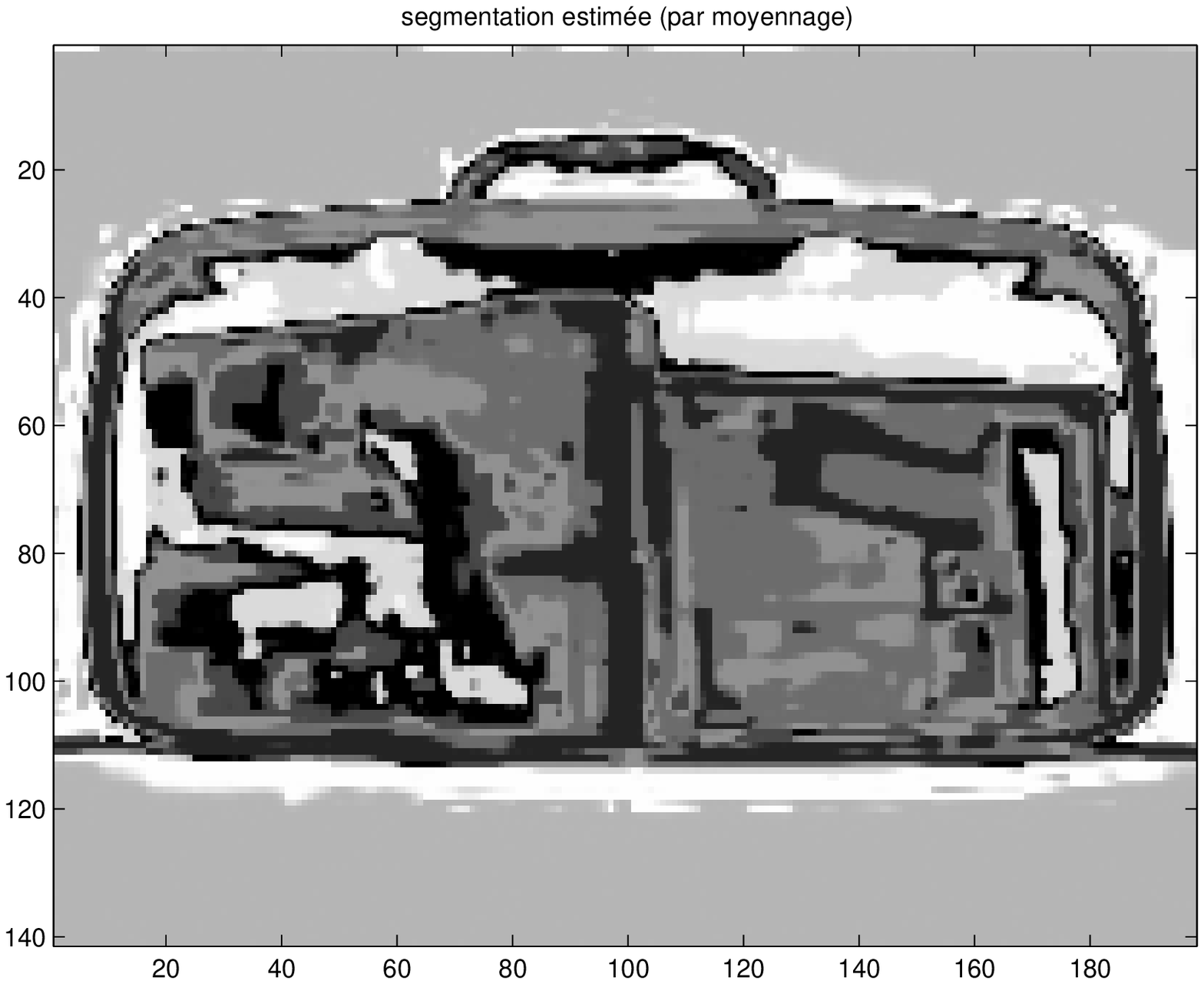,width=3.5 cm}} 
\\
 (a) & (b) & (c) & (d)
\end{tabular}

\begin{tabular}{ccc}
\mbox{\epsfig{file=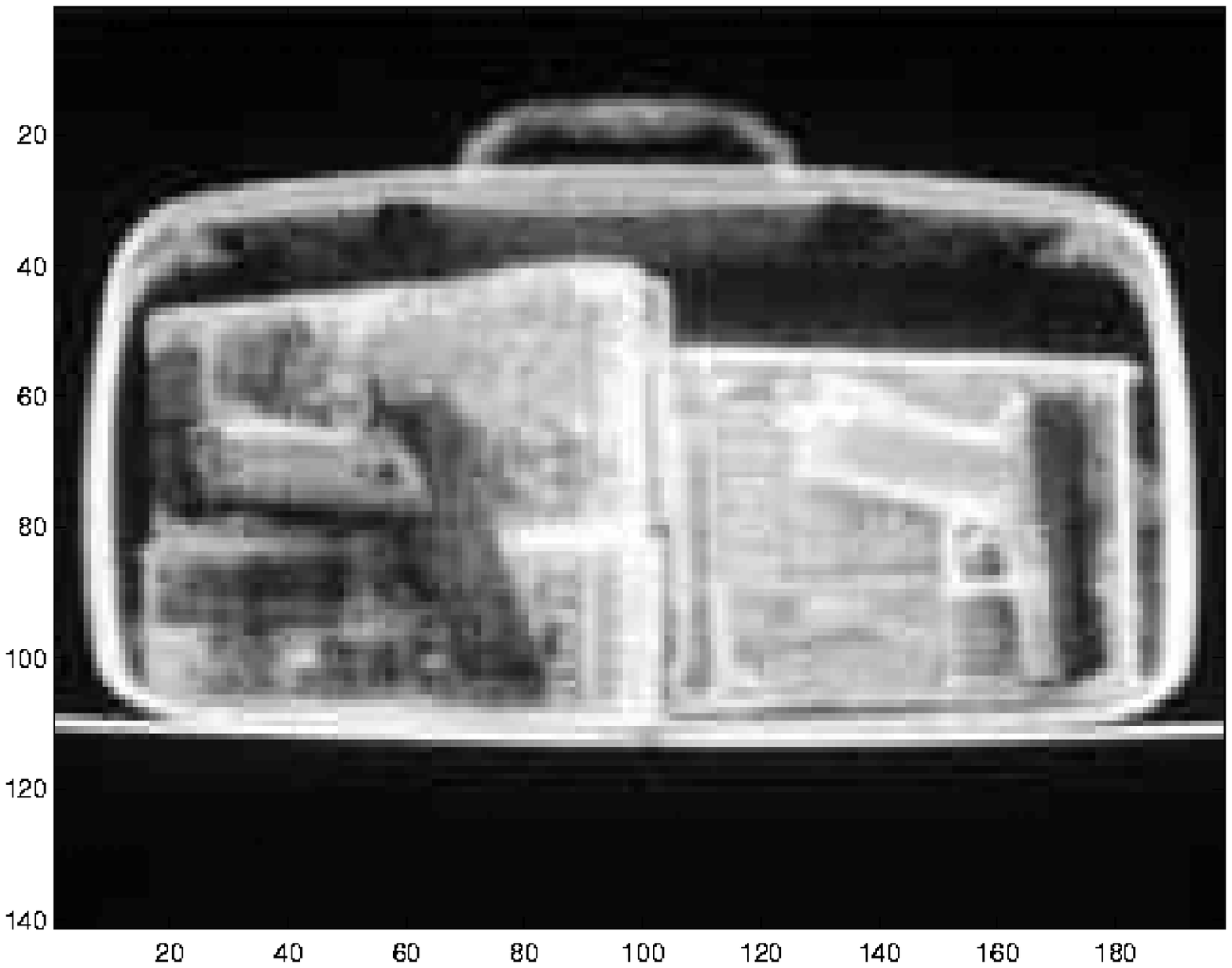,width=3.5 cm}} & 
\mbox{\epsfig{file=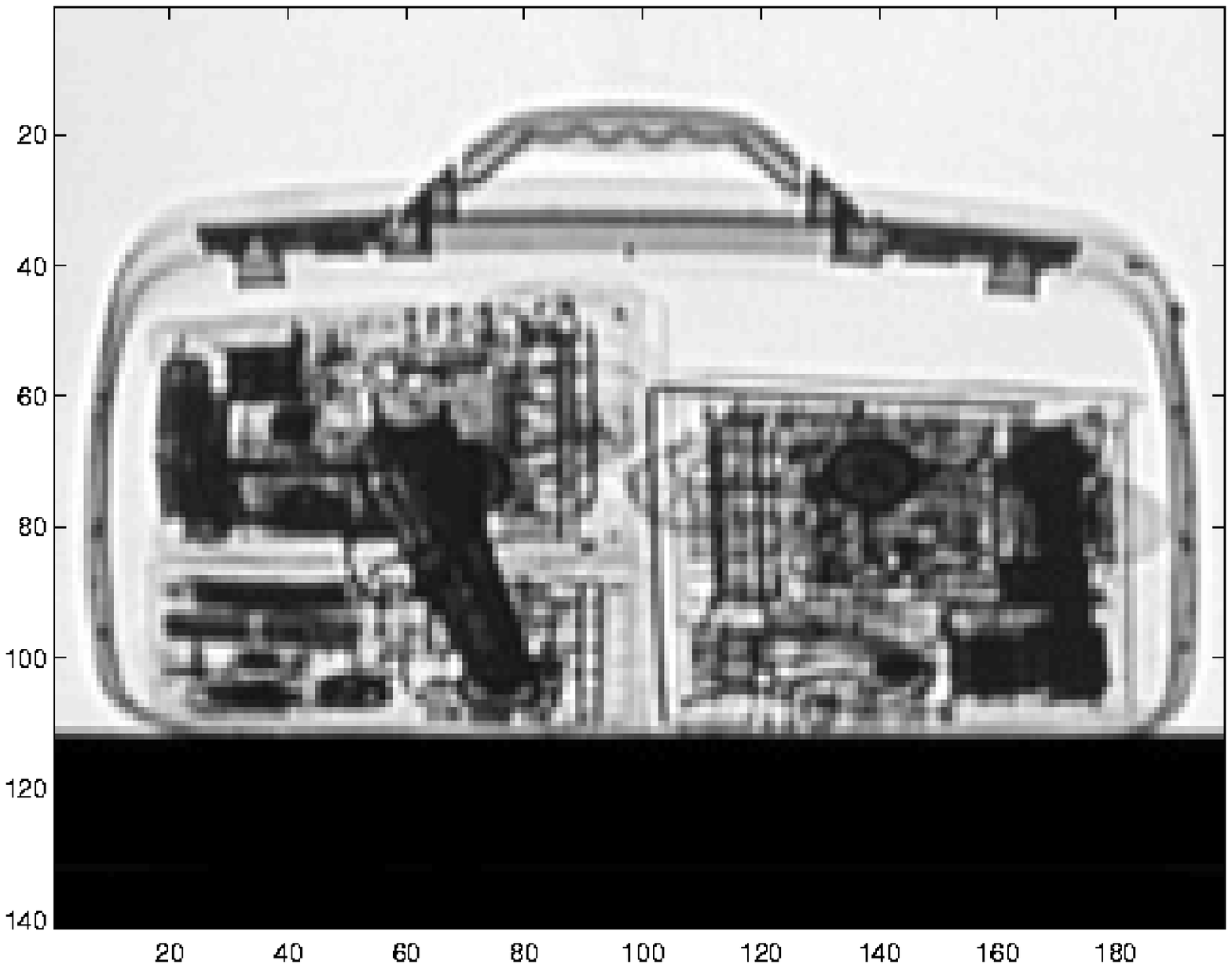,width=3.5 cm}} &	
\mbox{\epsfig{file=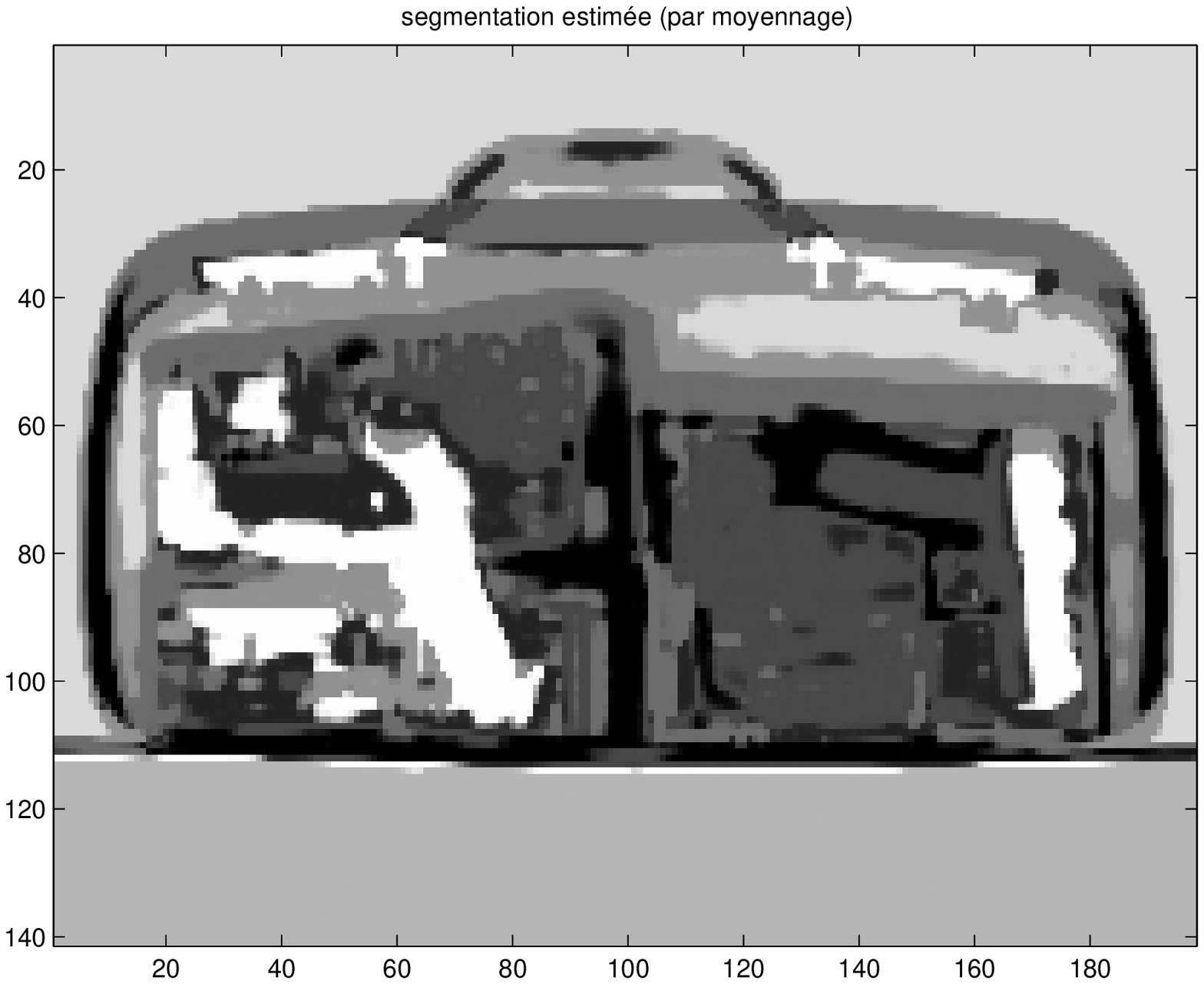,width=3.5 cm}} \\
	(e) & (f) & (g)
\end{tabular}
\caption{Results of data fusion from X ray images. a,b) two observations from transmission and backscattering X rays. c) segmentation of only image a) with 7 labels. d) segmentation of only image a)  with 8 labels. e,f) respective reconstruction of the images. g) result of data fusion with 8 labels.}
\label{valise}
\end{figure}
\noindent
In both applications we have satisfactoring results of image fusion, even when images present a great number of homogeneous regions and boundaries. Figures \ref{medical} and \ref{valise} show that the proposed method uses both images to increase the performances of the segmentation. Note also that the segmentation time of one image independantly or as result of the image fusion is practically the same. Indeed the proposed method does not really increase the complexity, making fusion and reconstruction in the same time. \\
\noindent
However in both cases the reconstructed images are not visibly improved. This is mostly due to the assumption that the values of $f_i(r)$ at any two different pixels are independant. We may expect for better results of reconstruction and segmentation if we introduce some local spatial dependancy between the neighboring pixels of images $f_i(r)$. This point is under developpment and we will report soon on the results.


\section{Conclusion}
\noindent
We proposed a Bayesian method for data fusion of images, with a Potts Markov Random Field model on the hidden variable $\zb$. We illustrated how the segmentation is improved by using data fusion through two applications : MRI and CT images in medical imaging and X ray transmission and backscattering images in security systems. \\
\noindent
We showed then how reconstruction and fusion can be computed in the same time using a MCMC algorithm, which reduce the complexity of the algorithm. 

\nocite{Gindi93,Gautier98,Hebert89,Geman92,Aubert97,Charbonnier97,Robert96}

\small 

\bibliographystyle{ieeetr}

\bibliography{revuedef,biben,baseAJ,baseKZ,gpipubli,amd99,IAPQR03}

\begin{thebibliography}{10}

\bibitem{Gautier95b}
S.~Gautier, G.~Le~Besnerais, A.~Mohammad-Djafari, and B.~Lavayssi\`ere, {\em
  Data fusion in the field of non destructive testing}.
\newblock {M}aximum {E}ntropy and {B}ayesian {M}ethods, Santa Fe, \sca{nm}:
  Kluwer Academic Publ., {K.~H}anson~ed., 1995.

\bibitem{Boyd94}
J.~Boyd and J.~Little, ``Complementary data fusion for limited-angle
  tomography,'' in {\em IEEE Proceeding of Computer vision and Pattern
  recognition}, (Seattle), pp.~288--294, 1994.

\bibitem{Boyd95}
J.~Boyd, ``Limited-angle computed tomography for sandwich structures using data
  fusion,'' in {\em journal of Nondestructive Evaluation, vol.14, no.2},
  pp.~61--76, 1995.

\bibitem{Matsopoulos}
G.~Matsopoulos, S.~Marshall, and J.~Brunt, ``Multiresolution morphological
  fuson of mr and ct images of the human brain,'' in {\em IEEE Proceedings on
  Vision, Image and Signal Processing, vol.141 Issue : 3}, (Seattle),
  pp.~137--142, 1994.

\bibitem{Bass00}
T.~Bass, ``Intrusion detection systems and multisensor data fusion,'' in {\em
  Comm. of the ACM, vol. 43}, pp.~99--105, 2000.

\bibitem{Snoussi02c}
H.~Snoussi and A.~Mohammad-Djafari, ``{I}nformation {G}eometry and {P}rior
  {S}election.,'' in {\em Bayesian Inference and Maximum Entropy Methods}
  (C.~Williams, ed.), pp.~307--327, MaxEnt Workshops, {A}merican {I}nstitute of
  {P}hysics, {A}ugust 2002.

\bibitem{Gindi93}
G.~Gindi, M.~Lee, A.~Rangarajan, and I.~G. Zubal, ``{B}ayesian reconstruction
  of functional images using anatomical information as priors,'' {\em
  \uppercase{ieee} {T}ransactions on {M}edical {I}maging}, vol.~12, no.~4,
  pp.~670--680, 1993.

\bibitem{Gautier98}
S.~Gautier, J.~Idier, A.~Mohammad-Djafari, and B.~Lavayssi\`ere, ``{X}-ray and
  ultrasound data fusion,'' in {\em {P}roceedings of the {I}nternational
  {C}onference on {I}mage {P}rocessing}, (Chicago, \sca{il}), pp.~366--369,
  {O}ctober 1998.

\bibitem{Hebert89}
T.~Hebert and R.~Leahy, ``A generalized \sca{em} algorithm for 3-{D} {B}ayesian
  reconstruction from {P}oisson data using {G}ibbs priors,'' {\em
  \uppercase{ieee} {T}ransactions on {M}edical {I}maging}, vol.~8,
  pp.~194--202, {J}une 1989.

\bibitem{Geman92}
D.~Geman and G.~Reynolds, ``Constrained restoration and the recovery of
  discontinuities,'' {\em \uppercase{ieee} {T}ransactions on {P}attern
  {A}nalysis and {M}achine {I}ntelligence}, vol.~14, pp.~367--383, {M}arch
  1992.

\bibitem{Aubert97}
G.~Aubert and L.~Vese, ``A variational method in image recovery,'' {\em
  \uppercase{siam} {J}ournal of {N}umerical {A}nalysis}, vol.~34,
  pp.~1948--1979, {O}ctober 1997.

\bibitem{Charbonnier97}
P.~Charbonnier, L.~Blanc-F\'eraud, G.~Aubert, and M.~Barlaud,
  ``De{\-}ter{\-}ministic edge-preserving regularization in computed imaging,''
  {\em \uppercase{ieee} {T}ransactions on {I}mage {P}rocessing}, vol.~6,
  pp.~298--311, {F}ebruary 1997.

\bibitem{Robert96}
C.~Robert, {\em M\'ethodes de {M}onte-{C}arlo par cha\^{\i}nes de {M}arkov}.
\newblock Paris, France: Economica, 1996.

\end{thebibliography}

\edoc